\documentclass[aps,prb,twocolumn,superscriptaddress,floatfix,letterpaper]{revtex4-1}

\pdfoutput=1

\usepackage[bbgreekl]{mathbbol}
\usepackage{amscd}

\begin{document}

\title{
Modular transformation and bosonic/fermionic topological orders\\
 in Abelian fractional quantum Hall states
}

\author{Xiao-Gang Wen}
\affiliation{Perimeter Institute for Theoretical Physics, Waterloo, Ontario, N2L 2Y5 Canada}
\affiliation{Department of Physics, Massachusetts Institute of
Technology, Cambridge, Massachusetts 02139, USA}
\affiliation{Institute for Advanced Study, Tsinghua University,
Beijing, 100084, P. R. China}

\begin{abstract}


The non-Abelian geometric phases of the robust degenerate ground states were
proposed as physically measurable defining properties of topological order in
1990.  In this paper we discuss in detail such a quantitative characterization
of topological order, using generic Abelian fractional quantum Hall states as
examples.  We find a way to determine even the pure $U(1)$ phase of the
non-Abelian geometric phases.  We show that the non-Abelian geometric phases
not only contain information about the quasi-particle statistics, they also
contain information about the chiral central charge $\Del c$ mod 24 of the edge
states, as well as the Hall viscosity.  Thus, the non-Abelian geometric phases
(both the Abelian part and the non-Abelian part) provide a quite complete
way to characterize 2D topological order. 


\end{abstract}

\maketitle

{\small \setcounter{tocdepth}{2} \tableofcontents }

\section{Introduction}

Topological order\cite{Wrig,WNtop} is a new kind of order that is beyond Landau
symmetry breaking theory,\cite{L3726,LanL58}  and appear in fractional quantum
Hall (FQH) states.\cite{TSG8259,L8395} It represents the pattern of long range
entanglement in highly entangled quantum ground
states.\cite{KP0604,LW0605,CGW1038}  The notion of topological order was first
introduced (or partially defined) via a physically measurable property: the
topologically robust ground state degeneracies (\ie robust against any local
perturbations that can break all the symmetries).\cite{WNtop} The ground state
degeneracies depend on the topology of the space.  Shortly after, a more
complete characterization/definition of topological order was introduced
through the topologically robust non-Abelian geometric phases associated with
the degenerate ground states as we deform the Hamiltonian.\cite{Wrig,KW9327} It
is was conjectured that the non-Abelian geometric phases may provide a complete
characterization of topological orders in 2D,\cite{Wrig} and, for example, can
provide information on the quasi-partiicle statistics. Recently, such
non-Abelian geometric phases were calculated
numerically,\cite{ZGT1251,CV1223,ZMP1233} and the connection to
quasi-partiicle statistics was confirmed.

We like to stress that, now we know many ways to characterize topological
orders, such as $K$-matrix,\cite{WNtop,BW9045,R9002,WZ9290,FS9333}
$F$-symbol,\cite{LW0510} fixed-point tensor network,\cite{GLS0918,BAV0919}
\etc. But they are either not directly measurable (\ie correspond to a
many-to-one characterization), or cannot describe all topological orders, or
cannot fully characterize topological orders (\ie correspond to a one-to-many
characterization). So the non-Abelian geometric phases, as a physical and
potentially complete characterization of topological orders, are very
important.  To understand this point, it is useful to compare topological order
with superconducting order.  Superconducting order is defined by physically
measurable defining properties: zero resistivity and Meissner effect.
Similarly topological order is defined by physically measurable defining
properties: topological ground state degeneracy and the non-Abelian geometric
phases of the degenerate ground states.

To gain a better understanding of non-Abelian geometric phases of the
degenerate ground states, in this paper, we are going to study the non-Abelian
geometric phases of most general Abelian fractional quantum Hall (FQH) states
in detail.  In particular we find a way to determine both the non-Abelian part
and the Abelian $U(1)$ phase of the non-Abelian geometric phases.  Through
those examples, we like to argue that non-Abelian geometric phases form a
representation of $SL(2,\Z)$, which is generated by $S,T$ satisfying
$S^4=(ST)^6=1$.  Such a  representation contains information on the
quasi-particle statistics as well as the chrial central charge $\Del c$ mod 24.

It is generally believed that the quasi-particle statistics and the chiral
central charge completely determine the topological order.  Thus the
non-Abelian geometric phases provide a quite complete way to determine the
topological order.  In this paper, we will calculate and study the non-Abelian
geometric phases for a generic Abelian FQH state.  First, we will give a
general introduction of topological order and FQH states.

\section{Topological orders in FQH states}

\subsection{Introduction to FQH effect}

At high enough temperatures, all matter is in a form of gas.  However, as
temperature is lowered, the motion of atoms becomes more and more correlated.
Eventually the atoms form a very regular pattern and a crystal order is
developed.  
With advances of semiconductor technology, physicists learned how to confine
electrons on a interface of two different semiconductors, and hence making a
two dimensional electron gas (2DEG).  In early 1980's physicists put a 2DEG
under strong magnetic fields ($\sim$ 10 Tesla) and cool it to very low
temperatures ($\sim 1K^\circ$), hoping to find a crystal
formed by electrons. They failed to find the electron crystal. But
they found something much better. They found that the 2DEG forms a new kind of
state -- Fractional Quantum Hall (FQH) state.\cite{KDP8094,TSG8259} Since the temperatures
are low and interaction between electrons are strong, the FQH state is a
strongly correlated state. However such a strongly correlated state is not a
crystal as people originally expected. It turns out that the strong quantum
fluctuations of electrons due to their very light mass prevent the formation
of a crystal.  Thus the FQH state is a quantum liquid.\footnote{A crystal can
be melted in two ways: (a) by thermal fluctuations as we raise temperatures
which leads to an ordinary liquid; (b) by quantum fluctuations as we reduce the
mass of the particles which leads to a quantum liquid.}


Soon many amazing properties of FQH liquids were discovered.  A FQH liquid is
more ``rigid'' than a solid (a crystal), in the sense that a FQH liquid cannot
be compressed. Thus a FQH liquid has a fixed and well-defined density.  More
magic shows up when we measure the electron density in terms of filling factor
$\nu$, defined as a ratio of the electron density $n$ and the applied magnetic
field $B$
\[
\nu\equiv \frac{n h c}{e B} 
=\frac{\hbox{density of electrons}}{\hbox{density of magnetic flux quanta}}
\]
All discovered FQH states have such densities
that the filling factors are exactly given by some rational numbers,
such as $\nu=1, 1/3, 2/3,2/5, ...$.
FQH states with simple rational filling factors 
(such as $\nu=1, 1/3, 2/3,...$) are more stable and easier to observe.
The exactness of the filling factor leads to an exact Hall resistance
$\rho_{xy}=\nu^{-1}\frac{h}{e^2}$.
Historically, it was the discovery of exactly quantized Hall resistance that
led to the discovery of quantum Hall states.\cite{KDP8094}  Now the exact Hall
resistance is used as a standard for resistance.

\subsection{Dancing pattern -- a correlated quantum fluctuation}

Typically one thinks that only crystals have internal pattern, and liquids 
have no internal structure.
However, knowing FQH liquids with those magical filling factors, one cannot help
to guess that those FQH liquids are not ordinary featureless liquids.
They should have some
internal orders or ``patterns''. According to this picture,
different magical filling factors can be thought to come from different 
internal ``patterns''. The hypothesis of
internal ``patterns'' can also help to explain the ``rigidness'' ({\it ie.}
the incompressibility) of FQH liquids. Somehow, a compression of
a FQH liquid could break its internal ``pattern'' and cost a finite energy.
However, the hypothesis of internal ``patterns'' appears to have one difficulty
-- FQH states are liquids, and how can liquids have internal ``patterns''?

Theoretical studies since 1989 did conclude that FQH liquids contain highly
non-trivial internal orders (or internal ``patterns''). However these internal
orders are different from any other known orders and cannot be characterized by
local order parameters and long range correlations.  What is really new (and
strange) about the orders in FQH liquids is that they are not associated with
any symmetries (or the breaking of symmetries),\cite{Wrig,WNtop} and
cannot be characterized by order parameters associated with broken symmetries.
This new kind of order is called {\it topological order},\cite{Wrig,WNtop}
and we need a whole new theory to describe the topological orders.

\begin{figure}
\centerline{
\includegraphics[scale=0.2]{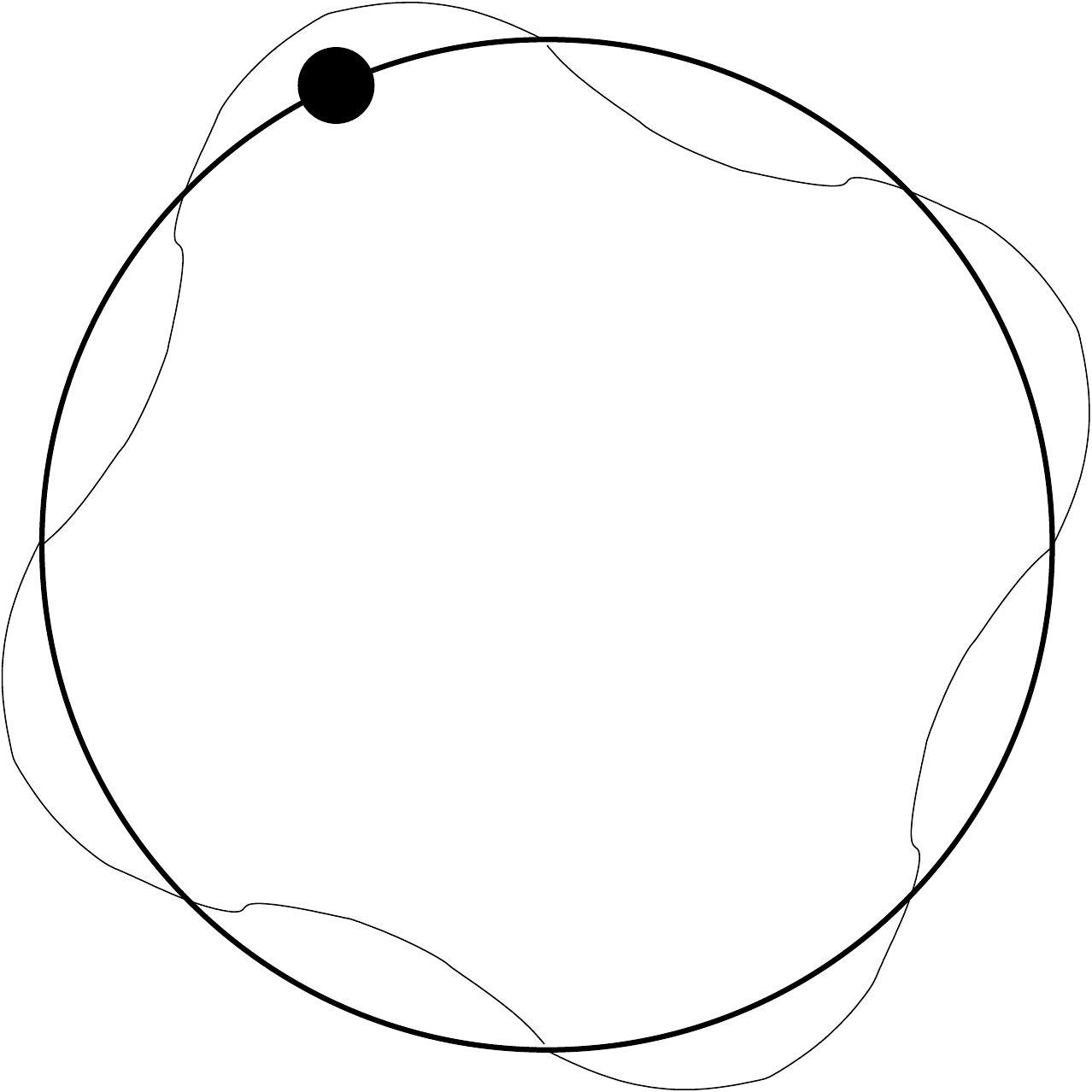}
}
\caption{
A particle and its quantum wave on a circle.
}
\label{topartf1}
\end{figure}

To gain some intuitive understanding of those internal ``patterns'' (\ie
topological orders), let us try
to visualize the quantum motion of electrons in a FQH state. 
Let us first consider a single electron in a magnetic field.  Under the
influence of the magnetic field, the electron always moves along circles (which
are called cyclotron motions).  In quantum physics, only certain discrete
cyclotron motions are allowed due the wave property of the particle.  The
quantization condition is such that the circular orbit of  an allowed cyclotron
motion contains an integer number of wavelength (see Fig.  \ref{topartf1}). The
above quantization condition can be expressed in a more pictorial way. We may
say that the electron dances around the circle in steps.  The step length is
always exactly equal to the wavelength. The quantization condition requires
that the electron always takes an integer number of steps to go around the
circle.  If the electron takes $m$ steps around the circle we say that the
electron is in the $(m+1)^\text{th}$ Landau level.  ($m=0$ for the first Landau
level.) Such a motion of an electron give it an angular momentum $m \hbar$.  We
will call such an angular momentum (in unit of $\hbar$) the orbital spin
$S_{ob}$ of the electron. So $S_{ob}=m$ for electrons in the $(m+1)^\text{th}$
Landau level.

The orbital spin can be probed via its coupling to the curvature of the 2D
space.  As an object with no-zero orbital spin moves along a loop $C$ on a
curved 2D space, a non-zero geometric phase will be induced. The geometric phase is
given by
\begin{align}
\phi =S_{ob} \oint_C d x^i \om_i =S_{ob} \int_{S_C} d^2 x R
\end{align}
where $S_C$ is the area enclosed by the loop $C$. Here $R$ is the the Gauss
curvature and $\om_i$ the connection whose curl gives rise to the curvature.
The relation between $\om_i$ and $R=\veps_{ij}\prt_i \om_j$ is the same as
relation between the gauge potential $a_i$ and the ``magnetic field"
$b=\veps_{ij}\prt_i a_j$.  On a sphere, we have $\int_\text{sphere} d^2 x R
=4\pi$.


When we have many electrons to form a 2DEG, electrons not only do their own
cyclotron dance in the first Landau level, they also go around each other and
exchange places. Those additional motions also subject to the quantization
condition. For example an electron must take an integer steps to go around
another electron.  Since electrons are fermions,  exchanging two electrons
induces a minus sign to the wave function. Also exchanging two electrons is
equivalent to move one electron half way around the other electron. Thus an
electron must take half integer steps to go half way around another electron.
(The half integer steps induce a minus sign in the electron wave.) In other
words an electron must take an odd-integer steps to go around another
electron.  Electrons in a FQH state not only move in a way that satisfies the
quantization condition, they also try to stay away from each other as much as
possible, due the strong Coulomb repulsion and the Fermi statistics between
electrons. This means that an electron tries to take more steps to go around
another electron.

Now we see that the quantum motions of electrons in a FQH state 
are highly organized.
All the electrons in a FQH state dance collectively
following strict dancing rules: 
\begin{enumerate}
\item
all electrons do their own cyclotron dance in the $m^\text{th}$ Landau level
(\ie $m$ step around the circle); 
\item 
an electron always takes odd integer steps to go around another electron; 
\item
electrons try to stay away from each other (try to take more steps to go
around other electrons). 
\end{enumerate}
If every electrons follows these strict dancing rules, then only one unique
global dancing pattern is allowed. Such a dancing pattern describes the
internal quantum motion in the FQH state. It is this global dancing pattern
that corresponds to the topological order in the FQH
state.\cite{Wrig,WNtop}  Different QH states are distinguished by their
different dancing patterns (or equivalently, by their different topological
orders).

Recently, it was realized that topological orders are nothing but patterns of
long range entanglements.\cite{CGW1038} FQH states are highly entangled state
with long range entanglements. The dancing pattern described above is a simple
intuitive way to described the patterns of long range entanglements in FQH
states.

\subsection{Ideal FQH state and ideal FQH Hamiltonian}

A more precise mathematical description of the quantum motion of electrons
described above is given by the famous Laughlin wave function \cite{L8395}
\begin{equation}
\label{LauWF}
\Psi_m= \left[ \prod (z_i-z_j)^{m} \right] \e^{-\frac{1}{ 4} \sum |z_i|^2}
\end{equation}
where $z_j=x_j+i y_j$ is the coordinate of the $j^\text{th}$ electron.  The wave
function is antisymmetric when $m=$odd integer and correspond to a wave
function of fermions.  Such a wave function describes a filling factor
$\nu=1/m$ FQH state for fermionic electrons.  In this paper, we will also
consider the possibility of bosonic electrons.  A $\nu=1/m$ FQH state for
bosonic electrons is described by \eqn{LauWF} with $m=$ even.

We see that the wave function changes its phase by $2\pi m$ as we move one
electron around another. Thus an electron always takes $m$ steps to go around
another electron in the Laughlin state $\Psi_m$.  We also see that, as $z_i
\to z_j$, the wave function vanishes as $(z_i-z_j)^m$, so that the electrons
do not like to stay close to each other.  Since every electron takes the same
number of steps around another electron, every electron in the Laughlin state
is equally happy to be away from any other electrons.

The Laughlin wave function has another nice property: it is
the exact ground state of the following $N$-electron interacting
Hamiltonian
\begin{equation}
\label{HFQH}
   H_m = - \sum_{i=1}^N \frac{1}{2} 
(\v \prt_{\v r_i} - i\v  A(\v r_i) )^2 +\sum_{i<j} V_m(\v r_i-\v r_j)
\end{equation}
where $\v r_i=(x_i,y_i)$ is the coordinate of the $i^\text{th}$ electron and
\begin{equation}
\label{Vm}
V_m(\v r)= \sum_{l=1,\cdots,m-1} v_l (-)^l \prt^l_{z^*} \del(z) \prt^l_{z},\ \ \ \ \ 
z\equiv x+i y .
\end{equation}
To see $\Psi_m(\{z_i\})$ to be the ground state of the above Hamiltonian,
we note that all electrons in $\Psi_m(\{z_i\})$ are in the first Landau
level. Thus we can treat the kinetic energy $- \sum_{i=1}^N \frac{1}{2} 
(\v \prt_{\v r_i} - i\v  A(\v r_i) )^2$ as a constant.
We also note that average total potential energy of
$\Psi_m(\{z_i\})$ vanishes:
\begin{equation*}
V_{tot}\equiv
 \int \prod_id^2z_i\; 
\Psi_m^*(\{z_i\}) \sum_{i<j} V_m(z_i-z_j) \Psi_m(\{z_i\}) =0 .
\end{equation*}
This is because the function $\prt^l_{z_i-z_j}\Psi_m$ still have $m-l$ order
zero as $z_i\to z_j$. Thus $\del(z_i-z_j)\prt^l_{z_i-z_j}\Psi_m=0$ is $l<m$.
When $v_l>0$, the potential energy is positive definite.  Thus
$\Psi_m(\{z_i\})$ is the exact ground state of the Hamiltonian \eq{HFQH}.

If we compress the Laughlin state, some $m^\text{th}$ order zeros between pairs of
electrons become lower order zeros; so some electrons take less steps around
other electrons. This represents a break down of dancing pattern (or creation
of defects of topological orders), and cost finite energies.  Indeed, for such
a compressed wave function, the average total potential is greater than zero
since the order of zero in the compressed wave function can less or equal to
the $l$ in the potential \eq{Vm}.  Thus the Laughlin state explains the
incompressibility observed for $\nu=1/m$ FQH states.

Since the ground state wave function of the Hamiltonian $H_m$ \eq{HFQH} is
known exactly and describe a FQH state, we call $H_m$ an ideal FQH Hamiltonian.
Its exactly ground state $\Psi_m$ is called ideal FQH states.  From the ideal
FQH wave function $\Psi_m$, we can clearly see the dancing pattern of electrons.

\subsection{Pattern of zeros characterization of FQH states}

In the above, we have related the dancing patterns (or more generally, patterns
of long range entanglements) to the patterns of zeros in the ideal FQH wave
functions. Such an idea was developed further in \Ref{WW0808}.
(see also \Ref{SL0604,BKW0608,SY0802,BH0802,BH0802a,BH0882}.)

A set of data called the patterns of zeros\cite{WW0808} was introduced to
characterize the  dancing patterns (or pattern of long range entanglements). By
definition, the pattern of zeros is a sequence of integers $\{S_a,
a=2,3,4,...\}$, where $S_a$ is the lowest order of zeros in the ideal wave
function as we fuse $a$ electrons together.  Naturally a pattern of zeros must
satisfy certain consistent conditions so that it does correspond to an existing
FQH wave functions.  The pattern of zeros that satisfy the consistent
conditions can be divided into two classes: quasi-periodic solutions with a
period $n$ and non quasi-periodic solutions.  The non quasi-periodic solutions
may not correspond to any gapped FQH states, so we only consider the
quasi-periodic solutions.  Those solution is said to satisfy the $n$-cluster
condition.  This quasi-periodicity reduces the infinite sequence $\{S_a,
a=2,3,4,...\}$ into a finite set of data $\{n;m;S_2,\cdots,S_n\}$.

It turns out that the pattern of zeros data $\{n;m;S_2,\cdots,S_n\}$ can
characterize many non-Abelian FQH states.\cite{MR9162,Wnab} We can calculate
the filling fraction, the ground state degeneracy, the number of types of
quasi-particles, the quasi-particle charges, the quasi-particles fusion algebra,
\etc from the data
$\{n;m;S_2,\cdots,S_n\}$.\cite{WW0809,BW0932,BW1001a,LWW1024} However, in this
paper, we will concentrate on Abelian FQH states, which can be characterized by
a simpler $K$-matrix.

\subsection{The $(K,\v q, \v s)$ characterization of general Abelian FQH states}

Laughlin states with filling fraction $\nu=1/m$ represent the simplest QH
states (or simplest topological orders).  
FQH states also appear at filling fractions $\nu=2/3, 2/5, 3/7,...$.  So
what kind of dancing pattern (or topological order) give rise to those QH
states? 

$\nu=1/m$ Laughlin states as simplest FQH states contain only one
component of incompressible fluid.  One way to understand certain more general
FQH states (called Abelian FQH states) is to assume that those states contain
several components of incompressible fluid.  The filling fraction $\nu=2/3,
2/5, 3/7,...$ FQH states
are all Abelian FQH states
and can be understood this way.

The dancing patterns (or the topological orders) in an Abelian FQH state can be
described in a similar way by the dancing steps if there are several types of
electrons in the state (such as electrons with different spins or in different
layers).  Each type of electrons form a component of incompressible fluid and
the Abelian FQH state contain several components of incompressible fluid.  It
turns out that even for an FQH state formed by a single kind of electrons, the
electrons can spontaneous separate into several ``different types'' and the FQH
state can be treated as a state formed by several different types of
electrons.\cite{BW9033,R9002}

With such a understanding, we find that the dancing patterns in an Abelian FQH
state can be characterized by an integer symmetric matrix $K$ and an integer
charge vector $\v q$.\cite{BW9033,R9002,FK9169,WZ9290} The dimension of $K$,
$\v q$, $\v s$ is $\ka$ which corresponds to the number of  incompressible
fluids in the FQH state.  An entry of $K$, $K^{IJ}$, is the number of steps
taken by a particle in the $I^\text{th}$ component to go around a particle in the
$J^\text{th}$ component.  We note that when $I=J$, the order of zero $K_{II}$
depends on the statistics of electrons. For bosonic electrons, $K_{II}=$ even
and for fermionic electrons, $K_{II}=$ odd.

An entry of $\v q$, $q_I$, is the charge (in unit of $e$) carried by the
particles in the $I^\text{th}$ component of the incompressible fluid.  An entry of
$\v s$, $s_I$, describe how the $I^\text{th}$ component of the incompressible fluid
couple to the curvature of the 2D space.  It is roughly given by the orbital
spin of the particles in the $I^\text{th}$ component.  Or more precisely,  if the
particle in the $I^\text{th}$ component are in the $m^\text{th}$ Landau level
$s_I=m-1+\frac{K_{II}}{2}$.\cite{Wtoprev} The term $m-1$ is the  orbital spin
of the particles in the $I^\text{th}$ component.  There is also a ``quantum''
correction $\frac{K_{II}}{2}$ that is beyond semi-classical picture of
electron dancing step.

All physical properties associated with the topological orders can be
determined in term of $K$ and $\v q$. For example the filling factor is simply
given by $\nu=\v q^TK^{-1}\v q$.

In the $(K,\v q, \v s)$ characterization of QH states, the $\nu=1/m$ Laughlin
state is described by $K=m$, $\v q=1$, and $\v s=m/2$, while the $\nu=2/5$
Abelian state is described by $K=\bpm3&2\\ 2&3\epm$, $\v q=\bpm1\\ 1\epm$, and
$\v s=\bpm 3/2\\ 5/2\epm$.

The  topological order described by $(K,\v q)$ have a precise mathematical
realization in term of ideal multilayer FQH wave function:
\begin{align}
\label{PsiK}
 \Psi_K &=\Big\{
\prod_{I<J}
\prod_{i,j} (z^{I}_i-z^{J}_j)^{K_{IJ}}
\Big\}
\times
\nonumber\\
& \ \ \ \ 
\Big\{
\prod_{I}
\prod_{i<j} (z^{I}_i-z^{I}_j)^{K_{II}}
\Big\}
\e^{-\frac{B}{2}\sum_{i,I} [y^{I}_i]^2}
\end{align}
where $I,J=1,\cdots,\ka$ label $\ka$ different layers and $z^{I}_i$ is the
coordinate of the $i^\text{th}$ electron in the $I^\text{th}$ layer.  For such
a multilayer system, electrons in each layer naturally from a component of
incompressible fluid.  Since those electrons all carry unit charges, so
$q_i=1$.  The exponents (or the order of zeros) $K_{IJ}$ form the symmetric
integer matrix $K=(K_{IJ})$. The topological order in the FQH state \eq{PsiK}
is described by $K=(K_{IJ})$, $q_I=1$, and $s_I=K_{II}/2$.

We would like to point out that there exists a Hamiltonian (of a form similar
to that in \eqn{HFQH}) such that the wave function $\Psi_K$ is its exact ground
state.  Thus $\Psi_K$ is an ideal FQH state. The dancing pattern and its $(K,\v
q,\v s)$ characterization can be seen clearly from the ideal FQH wave function
$\Psi_K$.

It is instructive to compare FQH liquids with crystals.  FQH liquids are
similar to crystals in the sense that they both contain rich internal patterns
(or internal orders).  The main difference is that the patterns in the crystals
are static related to the positions of atoms, while the patterns in FQH liquids
are ``dynamic'' associated with the ways that electrons ``dance" around each
other.  (Such a ``dynamic'' dancing pattern corresponds to pattern long range
entanglements.\cite{CGW1038}) However, many of the same questions for crystal
orders (and more general symmetry breaking orders) can also be asked and should
be addressed for topological orders. 

\subsection{Chern-Simons effective theory for Abelian FQH states}

We know that the low energy effective theory for a symmetry breaking order is
given by Ginzburg-Landau theory. What is the low energy effective theory of
topological order? One of the most important application of the $(K,\v q, \v
s)$ characterization of Abelian FQH states is to use the data $(K,\v q, \v s)$
to write down the low energy effective theory of the Abelian FQH states.

There are two types of low energy effective theories for Abelian FQH states:
Ginzburg-Landau-Chern-Simons effective theory\cite{GM8752,ZHK8982,R8986} and
pure Chern-Simons effective theory\cite{WNtop,BW9033,WZ9290}.  The pure
Chern-Simons effective theory is more compact and more closely related to the
$(K,\v q, \v s)$ characterization.  The detailed derivations of pure Chern-Simons
effective theory has been given in \Ref{Wtoprev,Wen04} and I will not repeat
them here. We find that an Abelian FQH states characterized by
$(K,\v q, \v s)$ is described by the following Chern-Simons effective theory
\begin{align}
\label{CSeffKq}
 \cL &=
-\sum_{I,J} \frac{K_{IJ}}{4\pi} a_{I\mu}\prt_\nu a_{J\la}\eps^{\mu\nu\la}
\\
&\ \ \ 
-\sum_{I} \frac{q_I}{2\pi} A_{\mu}\prt_\nu a_{I\la}\eps^{\mu\nu\la} 
+\sum_{I} \frac{s_I}{2\pi} \om_i\prt_\nu a_{I\la}\eps^{i\nu\la} +\cdots
\nonumber 
\end{align}
where $\cdots$ represents higher derivative terms [such as the Maxwell term
$(\prt_\mu a_{I\nu} -\prt_\nu a_{I\mu})^2$].  It is important to note that the
scales of the gauge fields $a_{I\mu}$ are chosen that the gauge charges of the
gauge fields $a_{I\mu}$ are all quantized as integers. Thus we cannot redefine
the gauge fields $a_{I\mu} \to \sum_J U_{IJ} a_{J\mu}$ to diagonalize $K$.

The Chern-Simons effective theory allows us to calculation all the topological
properties of Abelian FQH state. The filling fraction is given by $\nu=\v
q^TK^{-1}\v q$. The quasi-particle excitations are labeled by integer vectors
$\v l$. A component of $\v l$, $l_I$, corresponds to the integer gauge charge
of $a_{I\mu}$. 
In fact, a quasi-particle labeled by $\v l$ is described by the Following
wave function
\begin{equation*}
\Psi_{K,\xi}=\prod_{i,I}(z_i^{I}-\xi)^{l_I} \Psi_K. 
\end{equation*}
The electric charge of the quasi-particle $\v l$ is given by
$Q=\v q^T K^{-1}\v l$ and the statistics is given by $\th=\pi \v l^TK^{-1}\v
l$.

To compare with some later discussions, we introduce another way to label the
quasi-particles: we use a rational vector $\v \al=K^{-1}\v l$ to label the
quasi-particle labeled by $\v l$ in the previous scheme. A quasi-particle
labeled by $\v \al$ carries an electric charge $Q=\v q^T\v\al$ and has a
statistics $\th=\pi\v\al^TK\v\al$.

We note that, in the new labeling scheme, when $\v\al$ is an integral vector
$\v\al=\v n$, the corresponding quasi-particle excitation carries an integral
electric charge $Q=\v q^T\v n$.  The statistics of the excitations also agrees
with the statistics of $\v q^T\v n$ electrons.  So such an excitation can be
viewed as a bound state of a integral number of electrons.  We may view
quasi-particles that differ by several electrons as equivalent. Thus $\v\al$ is
equivalent to $\v\al'=\v\al+\v n$.  The equivalent classes of $\v\al$ correspond
to different types of fractionalized excitations. They correspond to $\v\al$'s
that lie inside of the $\ka$ dimensional unit cube.  Since the number of
$\v\al$'s inside of the $\ka$ dimensional unit cube is $|\det(K)|$, so there
are $|\det(K)|-1$ different types of the fractionalized quasi-particles.  (Note
$\v\al=0$ corresponds to the trivial unfractionalized excitations which
correspond to bound states of electrons.)

\subsection{The $(K,\v q, \v s)$ characterization is not physical enough}

As we have stressed that the key to understand topological order is to find a
characterization of  topological order.
For the real Hamiltonian of a FQH system
\begin{equation}
\label{FQHCoul}
   H_\text{real} = - \sum_{i=1}^N \frac{1}{2} 
(\v \prt_{\v r_i} - i\v  A(\v r_i) )^2 +\sum_{i<j} \frac{e^2}{|\v r_i-\v r_j|}
\end{equation}
the ground state wave function is very complicated.  We can hardly see any
pattern from such a complicated wave function.

\begin{figure}
\centerline{
\includegraphics[scale=0.5]{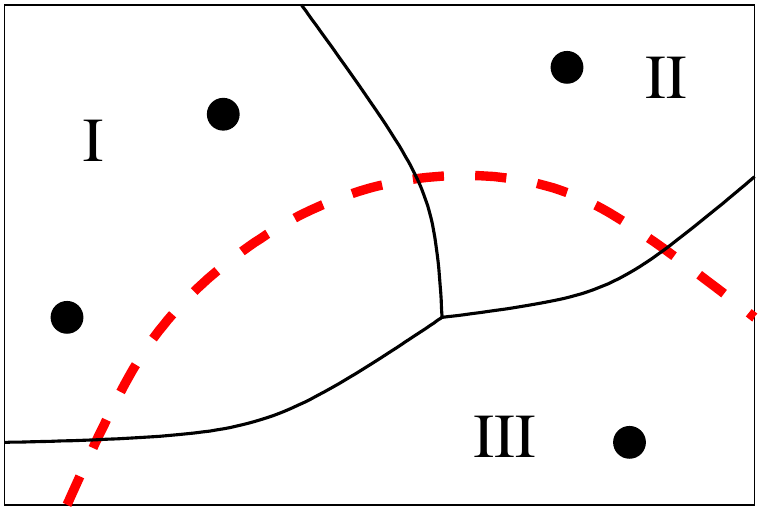}
}
\caption{(Color online)
The plane represents the space of possible Hamiltonians.
The solid lines are the phase transition lines that separate
phases with different topological orders.
The solid dots represent the ideal Hamiltonians whose ground state
can be characterized by $(K,\v q, \v s)$.
The dash line represents the real Hamiltonians that can be realized in
experiments.
}
\label{idlH}
\end{figure}

What we did above is to deform the Hamiltonian into an ideal Hamiltonian which
have a simple ideal ground state wave function.  We find that the ideal ground
state wave function can be characterized by $(K,\v q, \v s)$.  Then we say that all
the states in the same region of the phase diagram that contains the ideal
state are characterized by the same $(K,\v q, \v s)$ (see Fig. \ref{idlH}). This
allows us to characterize topological order in the ground state of real
Hamiltonian.

Such a understanding also indicates that the $(K,\v q,\v s)$ characterization
of topological order is not very direct. We cannot determine $K$ even if we
know the ground state wave function of a (non-ideal) Hamiltonian exactly.  In
other words, $K$ is not a directly measurable physical quantity.  We need to
rely on the deformation to ideal Hamiltonian to have a $(K,\v q, \v s)$
characterization of topological order.  This raises a possibility that two
different sets, $(K_1,\v q_1, \v s_1)$ and $(K_2,\v q_2, \v s_2)$, may
characterize the same topological order, since the two ideal FQH state
characterized by $(K_1,\v q_1, \v s_1)$ and $(K_2,\v q_2, \v s_2)$ may belong
to the same phase (see Fig. \ref{idlH}).  (\ie we can deform one ideal FQH
state into the other smoothly without encountering a phase transition.) Indeed,
through a field redefinition $a_{I\mu}\to \sum_j W_{IJ} a_{J\mu}$ in the
Chern-Simons effective theory \eq{CSeffKq}, we see that, if $(K_1,\v q_1,\v
s_1)$ and $(K_2,\v q_2,\v s_2)$ are related by\cite{Wtoprev}
\begin{equation}
\label{K2WK1W}
 W\v q_1=\v q_2,\ 
 W\v s_1=\v s_2,\  
W^TK_1W=K_2,\ 
W\in SL(\ka,Z),
\end{equation}
then $(K_1,\v q_1,\v s_1)$ and $(K_2,\v q_2,\v s_2)$ describe the same FQH
states.  Note that for $W\in SL(\ka,Z)$ the field redefinition $a_{I\mu}\to
\sum_j W_{IJ} a_{J\mu}$ does not change the quantization condition of the gauge
charges (\ie all gauge charges remain to be integers).

However, the equivalence condition \eq{K2WK1W} is not complete.  It was point
out in \Ref{H9590} that even $K$ matrices with different dimensions may
describe the same FQH state.\cite{KCW9963} Right now, we believe that all
Abelian FQH states are described by $(K,\v q,\v s)$. But $(K,\v q,\v s)$ is a
many-to-one labeling of Abelian topological order.  It is not clear if we have
found all the equivalence conditions of $(K,\v q,\v s)$'s.


The dancing pattern and the related $(K,\v q,\v s)$ description is an intuitive
way for us to visualize topological order (or pattern of long range
entanglements) in (Abelian) FQH states.  But, due to its many-to-one property,
$(K,\v q,\v s)$ is not something that can be directly measured in experiments
or numerical calculations.  So such a description of topological order,
although very pictorial, is not very physical.  A more rigorous and more
physical (partial) definition of topological order\cite{Wrig,WNtop} is
given by the topology-dependent ground state degeneracy which can be measured
in numerical calculations.  

In this paper, wen will concentrate on the physical characterization of
topological orders.  We will concentrate on the ground state degeneracy of FQH
states on compact spaces.  We obtain more physical data to characterized the
rich topological orders in FQH states, we will also study the non-Abelian geometric
phase associated with the degenerate  ground states as we deform the Hamiltonian
of the FQH systems.  First, to understand the ground state degeneracy of FQH
states, in the next section, we will study FQH state on torus.

\section{FQH states on torus}

\subsection{One electron in a uniform magnetic field}

One electron in a uniform magnetic field $B$ is described by the following
Hamiltonian
\begin{equation}
\label{HAxAy}
   H = - \frac{1}{2} \left [
(\frac{\partial}{\partial x} - i A_{x} )^2 +
( \frac{\partial}{\partial y} - i A_{y} )^2 \right ] ,
\end{equation}
where
\begin{equation}
\label{AxAy}
 (A_x,A_y)=(-By, 0) .
\end{equation}
The ground states of \eqn{HAxAy} are highly degenerate.
Those ground states form the first Landau level and have the following form
\begin{equation}
\label{psifz}
\Psi(x,y)=\Psi(z) = f( z) \e^{-\frac{B}{2} y^2}
\end{equation}
where the function $f(z)$ is a holomorphic
function of complex variables $z = x + iy$

First let us consider the symmetry of the Hamiltonian \eq{HAxAy}.
Since the magnetic field is uniform, we expect the translation symmetry
in both $x$- and $y$-directions.
The Hamiltonian \eq{HAxAy} does not depend on $x$, thus
\begin{equation*}
 T^\dag_{d_x\v x} H T_{d_x\v x} =H,\ \ \ \ \ \ 
T_{d_x \v x}=\e^{d_x \prt_x}
\end{equation*}
But the Hamiltonian \eq{HAxAy} depends on $y$ and there seems no translation
symmetry in $y$-direction.  However, we do have a translation symmetry in
$y$-direction ounce we include the gauge transformation.  
The Hamiltonian \eq{HAxAy} is invariant under $y\to y+d_y$ transformation
\emph{followed} by a $\e^{\imth d_y Bx+\imth \phi}$ $U(1)$-gauge transformation:
\begin{equation*}
T^\dag_{d_y\v y} H T_{d_y\v y} =H,\ \ \ \ \ \ 
T_{d_y \v y}=\e^{\imth d_y Bx+\imth \phi}\e^{d_y \prt_y} ,
\end{equation*}
In general
\begin{equation}
\label{TdH}
T^\dag_{\v d} H T_{\v d} =H,\ \ \ \ \ \ 
T_{\v d}=\e^{\imth d_y Bx+\imth \frac{B}{2}d_xd_y}\e^{\v d \cdot \v \prt} .
\end{equation}
where we have chosen the constant phase in $T_{\v d}$
\begin{equation}
\label{phiBdxdy}
 \phi=\frac{B}{2}d_xd_y 
\end{equation}
to simplify the later calculations.
The operator $T_{\v d}$ is called magnetic translation operator.  So the
Hamiltonian \eq{HAxAy} does have translation symmetry in any directions. But
the (magnetic) translations in different directions do not commute
\begin{equation}
\label{TdTd}
 T_{\v d_1} T_{\v d_2}
=\e^{\imth B (d_{1,x}d_{2,y}-d_{1,y}d_{2,x})} T_{\v d_2} T_{\v d_1}  .
\end{equation}
and momenta in $x$- and $y$-directions cannot be well defined at the same
time.

We note that $T_{\v d}$ acts on the total wave function
$f(z)\e^{-\frac{B}{2}y^2}$.  It is convenient to introduce
\begin{align*}
 \t T_{\v d}&= \e^{\frac{B}{2}y^2} T_{\v d} \e^{-\frac{B}{2}y^2}
\nonumber\\
 &= 
\e^{\imth d_y Bx+\imth \frac{B}{2}d_xd_y} 
\e^{\frac{B}{2}y^2} \e^{-\frac{B}{2}(y+d_y)^2}
\e^{\v d \cdot \v \prt}
\nonumber\\
 &= 
\e^{\imth d_y B (z +\frac{d_x+\imth d_y}{2})} \e^{\v d \cdot \v \prt}
\end{align*}
that act on the holomorphic part $f(z)$.

\subsection{One electron on a torus}

Now, let us try to put the above system on a torus by imposing the equivalence
or periodic conditions: $z\sim z+1$ and $z\sim z+\tau$ where
$\tau=\tau_x+\imth\tau_y$ is a complex number.  
%
%
%
Using the magnetic translation operator \eq{TdH}, we can put the
system on torus by requiring the wave functions to satisfy
\begin{equation}
\label{TdPsi}
 T_{(1,0)} \Psi(x,y)=\Psi(x,y),\ \ \ \ \ \
 T_{(\tau_x,\tau_y)} \Psi(x,y)=\Psi(x,y) ;
\end{equation}
\eqn{TdPsi} gives us
\begin{equation}
\label{qpconpsi}
\left\{ \begin{array}{l}
  \Psi (x + 1,y) = \Psi (x,y) , \\
  \Psi (x +\tau_x,y+\tau_y) = 
    \Psi (x, y) \e^{-\imth\tau_y B x -\imth \frac{B}{2} \tau_y\tau_x}\ ,
   \end{array} \right.
\end{equation}
which are called quasi-periodic conditions on
the wave-function.
In order for the above two conditions to be consistent,
the strength of magnetic field $B$ must satisfy
$\e^{-\imth\tau_y B (x+1)}=\e^{-\imth\tau_y B x}$ or
\begin{equation}
\label{BNphi}
B=2\pi \frac{N_\phi}{\tau_y}
\end{equation}
for an integer $N_\phi$. $N_\phi$ is the total number of flux quanta
going through the torus.  
\eqn{TdPsi} implies that
\begin{equation}
\label{Tdf}
\t T_{(1,0)} f(z)=f(z),\ \ \ \ \ \
\t T_{(\tau_x,\tau_y)} f(z)=f(z) ;
\end{equation}
or
\begin{equation*}
\left\{ 
\begin{array}{l}
  f(z+1) = f(z) \\
  f(z+\tau)=f(z) \e^{-\imth\tau \pi N_\phi-\imth 2\pi N_\phi z}\ .
   \end{array} \right.
\end{equation*}
The above can be written as
\begin{equation}
\label{qpcon}
  f(z+a+b\tau)=f(z) \e^{-\imth\tau \pi b^2 N_\phi-\imth 2\pi b N_\phi z}
,\ \ \ \ \ 
a,b=\text{ integers} . 
\end{equation}
The holomorphic functions that satisfy the above quasi-periodic conditions
have a form

\begin{equation}
\label{fzchin}
  f(z) = \sum_{n} \chi (n) \e^{\imth\frac{\pi\tau}{ N_\phi} n^2 + \imth 2 \pi n z} ,
\ \ \ \ \ \
\chi(n+N_\phi)=\chi(n) .
\end{equation}
We see that there are $N_\phi$ linearly independent holomorphic functions that
satisfy \eqn{qpcon}.  So there are $N_\phi$ states in the first Landau level on
a torus with $N_\phi$ flux quanta.
Note that
one can easily see that $f(z+1)=f(z)$.
To show $f(z+\tau)=f(z)\e^{-\imth\tau\pi N_\phi -\imth 2 \pi N_\phi z}$, we note that
\begin{align*}
f(z)&=\sum_{n} \chi (n) 
\e^{i\tau \frac{\pi}{ N_\phi} (n-N_\phi)^2 + \imth 2\pi (n-N_\phi) z}
\nonumber\\
&=
\e^{i\tau \pi N_\phi -\imth 2 \pi N_\phi z}
\sum_{n} \chi (n) 
\e^{ \frac{\imth\tau\pi}{ N_\phi} n^2 + \imth 2 \pi nz} \e^{-\imth 2 \pi\tau n }
\end{align*}
Therefore
\begin{align*}
f(z+\tau)&=
\e^{\imth\tau\pi N_\phi -\imth 2 \pi N_\phi (z+\tau)}
\sum_{n} \chi (n) 
\e^{\frac{\imth\tau\pi}{ N_\phi} n^2 + \imth 2 \pi n(z+\tau)} \e^{2\pi n }
\nonumber\\
&=
\e^{-\imth\tau\pi N_\phi -\imth 2 \pi N_\phi z}
\sum_{n} \chi (n) 
\e^{\frac{\imth\tau\pi}{ N_\phi} n^2 + \imth 2 \pi nz}
\end{align*}

A particular choice of the $N_\phi$ linearly independent holomorphic functions
can be obtained by choosing $\chi(n)$ in \eqn{fzchin} as
\begin{equation*}
 \chi(n)=1 \text{ if } n \text{ \% } N_\phi =l,\ \ \ \ \ \ \
 \chi(n)=0 \text{ if } n \text{ \% } N_\phi \neq l ,
\end{equation*}
where $a\% b \equiv a\text{ mod } b$, and $l=0,\cdots,N_\phi-1$.  We obtain
\begin{equation}
\label{thNphi}
 f^{(l)}(z|\tau)=\th_{l/N_\phi}(N_\phi z|N_\phi\tau)
=\sum_{n} \e^{ \imth\frac{\pi \tau}{N_\phi} (N_\phi n + l)^2 
+ \imth 2  \pi (N_\phi n +l) z },
\end{equation}
All holomorphic functions $f(z)$ that satisfy the condition \eq{qpcon} are
linear combinations of the above $N_\phi$ functions. 

\subsection{Translation symmetry on torus}

The one-electron system on the torus, \eqn{HAxAy}, has a translation symmetry
generated by the magnetic translation operator $T_{\v d}$ \eq{TdH}.  But to be
consistent with the quasi periodic condition \eq{TdPsi}, the allowed
translations must commute with $T_{(1,0)}$ and $T_{(\tau_x,\tau_y)}$.  
Those allowed translations are generated by two
elementary translation operators
\begin{equation}
\label{T1T2}
T_1=T_{(N_\phi^{-1},0)},\ \ \ \ \ 
T_2= T_{(\tau_x N_\phi^{-1},\tau_y N_\phi^{-1})}
\end{equation}
The corresponding operators that act on $f(z)$ are given by
\begin{equation}
\label{tT1tT2}
\t T_1= \t T_{(N_\phi^{-1},0)},\ \ \ \ \ 
\t T_2= \t T_{(\tau_x N_\phi^{-1},\tau_y N_\phi^{-1})}
\end{equation}
We note that
\begin{equation}
\label{Halg1}
T_1T_2 =\e^{\imth 2  \pi/N_\phi} T_2T_1, \ \ \ \ \ \ \ \
\t T_1\t T_2 =\e^{\imth 2  \pi/N_\phi} \t T_2\t T_1 .
\end{equation}
Such an algebra is called Heisenberg algebra.  Since $T_1$
and $T_2$ commute with the
Hamiltonian \eq{HAxAy}, the degenerate eigenstates of \eqn{HAxAy} form an
representation of the Heisenberg algebra.  
The Heisenberg algebra
has only one irreducible representation which is $N_\phi$ dimensional. Thus the
Hamiltonian \eq{HAxAy} must have $N_\phi$ fold degeneracy for each distinct
eigenvalue. Those $N_\phi$ fold degenerate states form a Landau level.  In the
first Landau level, those $N_\phi$ states are given by \eqn{thNphi} and they
satisfy
\begin{align}
\label{T1T2fl}
 \t T_1 f^{(l)}(z|\tau)&=
f^{(l)}(z+\frac1N_\phi|\tau)
=\e^{\imth 2  \pi l/N_\phi}  f^{(l)}(z|\tau)
\nonumber\\
 \t T_2 f^{(l)}(z|\tau)&=
f^{(l)}(z +\frac{\tau}{N_\phi}|\tau)
\e^{\imth \frac{\pi \tau}{N_\phi} +\imth 2  \pi z}
= f^{(l+1)}(z|\tau)
\end{align}

Note that from \eqn{thNphi}, we find
\begin{align*}
&\ \ \ f^{(l)}(z +\frac{\tau}{N_\phi}|\tau)
=\sum_{n} \e^{ \imth\frac{\pi \tau}{N_\phi} (N_\phi n + l)^2 
+ \imth 2  \pi (N_\phi n +l) (z+\frac{\tau}{N_\phi}) }
\nonumber\\
&=
\e^{-\imth \frac{\pi \tau}{N_\phi}} 
\sum_{n} \e^{ \imth\frac{\pi \tau}{N_\phi} (N_\phi n + l+1)^2 
+ \imth 2  \pi (N_\phi n +l) z }
\nonumber\\
&=f^{(l+1)}(z|\tau)
\e^{-\imth \frac{\pi \tau}{N_\phi} -\imth 2 \pi z} 
.
\end{align*}

\subsection{Many electrons on a torus}

A many electron system in a magnetic field is described by \eqn{HFQH}.
Assuming the magnetic field is given by
\eqn{BNphi} and 
If the filling fraction $\nu<1$ and the interaction is weak,
all the electrons will in the first Landau level. The many-electron wave
function will have a form
\begin{equation*}
 \Psi( \{x_i, y_i\})=f( \{z_i\}) \e^{-\frac{\pi N_\phi}{\tau_y} \sum y_i^2}
\end{equation*}
where we have assumed that the magnetic field is given by \eqn{BNphi}.
Here $z_i=x_i+\imth y_i$ is the coordinates for the $i^\text{th}$ electron. 

Assuming the space is a torus defined by
$z\sim z+1$ and $z\sim z+\tau$, the many-electron wave function
$\Psi( \{z_i\}) $
on torus must satisfy
\begin{equation}
\label{qpconpsiM}
\left\{ \begin{array}{l}
\Psi (\cdots,x_i + 1,y_i,\cdots) = \Psi (\cdots,x_i,y_i,\cdots) \\
\Psi (\cdots,x_i +\tau_x,y_i+\tau_y,\cdots) = 
\Psi (\cdots,x_i, y_i,\cdots) \times \\
\ \ \ \ \ \ \ \ \ \ \ \ \ \ \ \ \ \ \ \ \ \ \ \ \ \ \ \ \ \ \ \ \ \ \ \ \ \ \ \ 
 \e^{-\imth 2\pi N_\phi x_i -\imth \pi N_\phi \tau_x}\ .
   \end{array} \right.
\end{equation}
for each individual $i$.  This leads to a condition on $f(\{z_i\})$:
\begin{equation}
\label{qpconM}
f (\cdots,z_i +a+b\tau,\cdots) = 
f (\cdots, z_i,\cdots) \e^{-\imth \tau \pi b^2N_\phi -\imth 2\pi b N_\phi z_i} ,
\end{equation}
where $a,b=$ integers.

If the interaction potential between the electrons is give by $V_m$ in
\eqn{Vm}, than $\Psi(\{x_i, y_i\})$ is the exact zero energy eigenstate of
\eqn{HFQH} if $f(\{z_i\})$ has a $m^\text{th}$ order zero between
any pair of electrons:
\begin{equation}
\label{fmzero}
 f(\cdots, z_i,\cdots, z_j,\cdots) =|_{z_i\to z_j} O\big ( (z_i-z_j)^m \big ) .
\end{equation}

The holomorphic function $f( \{z_i\})$ that satisfy both \eqn{qpconM}
and \eqn{fmzero} exist only when the number of electrons $N$ satisfy
$N\leq N_\phi/m$. When $N=N_\phi/m$,
such $f( \{z_i\})$ has a form
\begin{align}
\label{ffcfr}
 f( \{z_i\})=f_c(Z) f_r( \{z_i\}),\ \ f_r( \{z_i\})=\prod_{i<j} 
\left[\th_{11}(z_i-z_j|\tau)\right]^m 
\end{align}
where $Z=\sum z_i$. The above $f( \{z_i\})$ can be viewed as the Laughlin wave
function $\prod_{i<j} (z_i-z_j)^m$ on torus 
[note that $\th_{11}(z_i-z_j|\tau)\sim z_i-z_j$ for small $z_i-z_j$].

From \eqn{qpcth11} we find
\begin{align*}
&\ \ \ f_r(\cdots, z_i+a+b\tau, \cdots)
\nonumber\\
&=
(-)^{m(N-1)(a+b)}
 f_r(\cdots, z_i, \cdots)
\nonumber\\&\ \ \ \
\e^{-\imth m \tau \pi b^2(N-1) - \imth 2m\pi  b [\sum_{j} (z_i - z_j) ] },
\nonumber\\
&=f_r(\cdots, z_i,\cdots) 
\nonumber\\&\ \ \ \
\e^{-\imth \tau \pi b^2N_\phi -\imth 2\pi b N_\phi z_i}
(-)^{m(N-1)(a+b)}
\e^{\imth m \tau \pi b^2 + \imth 2m\pi  b Z  }.
\end{align*}
We see that $f(\{z_i\})$ will satisfy the quasi periodic condition \eq{qpconM},
if $f_c(Z)$ satisfies
\begin{equation}
\label{qpconfc}
f_c(Z+a+b\tau)
=f_c(Z) (-)^{m(N-1)(a+b)}
\e^{-\imth m \tau \pi b^2 - \imth 2m\pi  b Z  }.
\end{equation}
There are $m$ linearly independent holomorphic functions
satisfying the above condition (see \eqn{fzchin}).
One particular choice of those $m$ linearly independent holomorphic functions
is given by (see \eqn{thlmab})
\begin{align}
\label{fclth}
 f_c^{(l)}(Z)&=\e^{\imth m\pi\frac{\tau}{4} + \imth m\pi (Z+\frac12)}
\th_{\frac{l}{m}}\left (m (Z+\frac12 +\frac\tau 2)\Big | m\tau \right ),
\nonumber\\ & 
\ \ \ \ \ \ \ \ \ \ \ \ \ \ \ \ \ \ \ \ \ \ \ \
\text{if }\ m(N-1)=\text{odd},
\nonumber\\
 f_c^{(l)}(Z)&=\th_{\frac{l}{m}}(mZ|m\tau),\ \ 
\text{if }\ m(N-1)=\text{even},
\end{align}
where $l=0,\cdots,m-1$.
Note that from \eqn{thlmab} we find that,
for $m(N_\phi-1)=$ even,
\begin{align*}
 f_c^{(l)}(Z+a+b \tau)&=
\th_{l/m}\left (m (Z+a+b\tau)\Big | m\tau \right )
\nonumber\\
& = \e^{-\imth\tau \pi m b^2 - \imth 2\pi Z b m} f_c^{(l)}(Z) ,
\end{align*}
for $m(N_\phi-1)=$ odd,
\begin{align*}
&\ \ \ \  f_c^{(l)}(Z+a+b \tau)
\nonumber\\
&= \e^{\imth m\pi\frac{\tau}{4} + \imth m\pi (Z+a+b\tau+\frac12)}
\th_{l/m}\left (m (Z+a+b\tau+\frac12 +\frac\tau 2)\Big | m\tau \right )
\nonumber\\
&=
\e^{\imth m\pi\frac{\tau}{4} + \imth m\pi (Z+a+b\tau+\frac12)}
\e^{-\imth\tau \pi m b^2 - \imth 2\pi (Z+\frac12 +\frac12 \tau) b m}\times
\nonumber\\
&\ \ \ \ \ \ \ 
\th_{l/m}\left (m (Z+\frac12 +\frac\tau 2)\Big | m\tau \right )
\nonumber\\
&=
\e^{\imth m\pi (a+b\tau)
-\imth\tau \pi m b^2 - \imth 2\pi (Z+\frac12 +\frac12 \tau) b m}
f_c^{(l)}(Z)
\nonumber\\
&= (-)^{a+b}\e^{-\imth\tau \pi m b^2 - \imth 2\pi Z b m} f_c^{(l)}(Z) ,
\end{align*}

We see that there are $m$ different ways to put the $\nu=1/m$ Laughlin wave
function $\prod_{i<j} (z_i-z_j)^m$ on a torus and keep the $m^\text{th}$ order
zeros.  Those $m$ wave functions on torus correspond to the $m$ exact ground
state of the Hamiltonian \eq{HFQH}.  Thus the $\nu=1/m$ Laughlin state 
has $m$-fold degeneracy on torus.

We like to stress that the $m^\text{th}$ order zero is a local condition (a
local dancing rule). From the above discussion, we see that such a local
condition leads to the $m$-fold ground state degeneracy on torus which is
global and topological property. So the discussion in this section represents
the rigorous mathematical theory that demonstrate how local dancing rules (the
$m^\text{th}$ order zero) determine the global dancing patterns and number of
global dancing patterns (the ground state degeneracy).  In general, if we want
to put a holomorphic function with $m^\text{th}$ order zeros on a genus $g$
Reimann surface, we will get $m^g$-fold degenerate ground states.

\subsection{Stability of ground state degeneracy}

For an $N$-electron system, the magnetic translation operators
that commute with the many-body Hamiltonian \eq{HFQH} are given by
\begin{equation}
\label{TdN}
T^{(N)}_{\v d}=\prod_i \e^{\imth d_y Bx_i+\imth \frac{B}{2} d_xd_y}
\e^{\v d \cdot \v \prt_i} .
\end{equation}
The corresponding $\t T_{\v d}$ operators that act on $f(\{z_i\})$ are given
by
\begin{equation}
\label{tTdN}
\t T^{(N)}_{\v d}=\prod_i \e^{\imth d_y B (z_i+ \frac{d_x+\imth d_y}{2})}
\e^{\v d \cdot \v \prt_i} .
\end{equation}
Again, on torus, the allowed magnetic translations are generated by
$\t T^{(N)}_1=\t T^{(N)}_{(N_\phi^{-1},0)}$ and
$\t T^{(N)}_2=\t T^{(N)}_{(\tau_x N_\phi^{-1},\tau_y N_\phi^{-1})}$.
$\t T^{(N)}_1$ and
$\t T^{(N)}_2$ generate
the following Heisenberg algebra
\begin{equation}
\label{HalgN}
\t T^{(N)}_1 \t T^{(N)}_2 =\e^{\imth 2  \pi N/N_\phi} \t T^{(N)}_2 \t T^{(N)}_1 
\end{equation}
So, if the filling fraction of the many-electron system is $\nu=N/N_\phi=1/m$,
the irreducible representation of the Heisenberg algebra will be $m$
dimensional.  The eigenstates of \eqn{HFQH} are all at least $m$-fold
degenerate, including the ground states. This is another way to show the
$m$-fold ground state degeneracy of the $\nu=1/m$ Laughlin state.

The $m$ ground states in \eqn{fclth} form a particular
irreducible representation of the Heisenberg algebra \eq{HalgN}.
Note that $\t T^{(N)}_{\v d}$ acts only on $f_c(Z)$.
When $m(N-1)=$ even, we have (see \eqn{T1T2fl} and \eqn{thNphi}):
\begin{align}
\label{T12flcE}
\t T^{(N)}_1 f^{(l)}_c(Z|\tau)&=
f^{(l)}_c(Z+\frac{N}{N_\phi}|\tau)
=\e^{\imth 2 \pi l/m} f^{(l)}_c(Z|\tau)
\nonumber \\
\t T^{(N)}_2
f^{(l)}_c(Z|\tau)&=\e^{ \imth 2 \pi Z +\imth \frac{\pi \tau N}{N_\phi}}
f^{(l)}_c(Z+\frac{N}{N_\phi}\tau|\tau)
\nonumber\\
&=
f^{(l+1)}_c(Z|\tau)
\end{align}
When $m(N-1)=$ odd, we have:
\begin{align}
\label{T12flcO}
\t T^{(N)}_1 f^{(l)}_c(Z|\tau)&=
f^{(l)}_c(Z+\frac{N}{N_\phi}|\tau)
=-\e^{\imth 2 \pi l/m} f^{(l)}_c(Z|\tau)
\nonumber \\
\t T^{(N)}_2
f^{(l)}_c(Z|\tau)&=\e^{ \imth 2 \pi Z +\imth \frac{\pi \tau N}{N_\phi}}
f^{(l)}_c(Z+\frac{N}{N_\phi}\tau|\tau)
\nonumber \\
&=-f^{(l+1)}_c(Z|\tau)
\end{align}
Note that
for $m(N-1)=$ odd, we have (see \eqn{thlmT1T2})
\begin{align*}
f_c^{(l)}(Z+\frac{1}{m})
&= \e^{\imth m\pi\frac{\tau}{4} + \imth m\pi (Z+\frac{1}{m}+\frac12)}
\times
\nonumber\\
&\ \ \ \ \ \ \
\th_{l/m}\left (m (Z+\frac{1}{m}+\frac12 +\frac\tau 2)\Big | m\tau \right )
\nonumber\\
&=-
\e^{\imth 2\pi l/m} f_c^{(l)}(Z)
\end{align*}
\begin{align*}
&\ \ \ \  f_c^{(l)}(Z+\frac{\tau}{m})
\nonumber\\
&= \e^{\imth m\pi\frac{\tau}{4} + \imth m\pi (Z+\frac{\tau}{m}+\frac12)}
\th_{l/m}\left (m (Z+\frac{\tau}{m}+\frac12 +\frac\tau 2)\Big | m\tau \right )
\nonumber\\
&=
\e^{\imth m\pi\frac{\tau}{4} + \imth m\pi (Z+\frac{\tau}{m}+\frac12)}
\e^{-\imth\tau \pi/m  - \imth 2\pi (Z+\frac12 +\frac12 \tau) }
\times
\nonumber\\
&\ \ \ \ \
\th_{(l+1)/m}\left (m (Z+\frac12 +\frac\tau 2)\Big | m\tau \right )
\nonumber\\
&=
\e^{\imth m\pi (\frac{\tau}{m})
-\imth\tau \pi/m  - \imth 2\pi (Z+\frac12 +\frac12 \tau)}
f_c^{(l+1)}(Z)
\nonumber\\
&= -\e^{-\imth\tau \pi/m - \imth 2\pi Z } f_c^{(l+1)}(Z) ,
\end{align*}

The eigenvalues of $\t T_1$ and $\t T_2$ are pure phases $\e^{\imth \vphi}$.  We
see that, in the large $N_\phi$ limit (with $N/N_\phi=1/m$ fixed), the phases
$\vphi$ are of order 1.  Since $\t T_1$ and $\t T_2$ are translation operators
by distances of order $1/N_\phi$, the momentum differences between different
degenerate ground states are of order $\Del k\sim N_\phi$.  The electron
density is of order $N_\phi$ and the separation between electrons is of order
$l_B\sim 1/\sqrt{N_\phi}$.  We see that $\Del k \gg 1/l_B$.

When we add impurities $\del H$ to break the translation symmetry, $\del H$
will split the $m$-fold ground state degeneracy of \eqn{HFQH} at $\nu=1/m$.
However, each $\del H$ can only cause a momentum transfer of order $1/l_B\sim
\sqrt{N_\phi}$. 
Thus in the first order perturbation theory,
$\del H$ cannot mix the degenerate ground state and cannot lift the degeneracy.
We need a $(\sqrt{N_\phi})^\text{th}$ order perturbation to
have a momentum transfer of order $\Del k\sim N_\phi$.  
Such a perturbation can only cause an energy splitting of order
\begin{equation*}
 \Del \sim u^{\sqrt{N_\phi}}=\e^{-L/\xi}
\end{equation*}
where $u$ is a dimensionless coupling constant to represent the strength of
the impurity Hamiltonian $\del H$, $L=\sqrt{N_\phi}$ represents the linear
size of the system, and $\xi=-\ln u$ represents a length scale.  We see that
the ground state degeneracy is robust against any random perturbation in the
thermodynamical limit.  The larger the system, the better the degeneracy.

We can also show that, on genus $g$ Reimann surface, the ground state
degeneracy of the $\nu=1/m$ Laughlin state is $m^g$.\cite{WNtop}  Such a
ground state degeneracy is also robust against any random perturbations.  The
existence of the robust topology-dependent ground state degeneracies implies
the existence of topological order.

\section{Generic Abelian FQH states 
}

\subsection{Generic Abelian FQH wave functions on torus}

We have seen that the topological order in the Laughlin state
$\prod (z_i-z_j)^m\e^{-\frac{B}{2}\sum y_i^2}$ 
can be probed by putting the state on torus,
which results in $m$-fold degenerate ground states.
A more generic topological order labeled by $K$ and $\v q^T=(1,\cdots,1)$
is described a multilayer wave function \eq{PsiK}.
Such a topological order can also be probed by putting
the wave function \eq{PsiK} on torus.

In terms of the theta function, the multilayer wave function \eq{PsiK}
becomes (see \eqn{ffcfr})
\begin{align}
\label{PsiKth}
 \Psi_K &= f(\{z^{I}_i\}) \e^{-\frac{B}{2}\sum_{i,I} [y^{I}_i]^2}
\\
f(\{z^{I}_i\})&= f_c(\{Z^{I}\}) f_r(\{z^{I}_i\})
\nonumber\\
f_r(\{z^{I}_i\}) &=
\Big\{
\prod_{I<J}
\prod_{i,j} \th^{K_{IJ}}_{11}(z^{I}_i-z^{J}_j|\tau)
\Big\}
\times
\nonumber\\
&\ \ \ \ \ 
\Big\{
\prod_{I}
\prod_{i<j} \th^{K_{II}}_{11}(z^{I}_i-z^{I}_j|\tau)
\Big\}
\nonumber 
\end{align}
on a torus. Here
\begin{align}
\label{ZIzI}
 Z^{I}=\sum_i z^{I}_i  
\end{align}
are the center-of-mass coordinates and
$f(\{z^{I}_i\})$ satisfies a quasi-periodic boundary condition
\begin{align}
\label{qpconMM}
& \ \ \  f (\cdots,z^{I}_i +a+b\tau,\cdots) 
\nonumber\\
& = 
f (\cdots, z^{I}_i,\cdots) 
\e^{-\imth \tau \pi b^2N_\phi -\imth 2\pi b N_\phi z^{I}_i} .
\end{align}
where $N_\phi$ is the total number of the magnetic flux quanta
through the torus.
From \eqn{qpcth11}, we find
\begin{align}
\label{frabM1}
&\ \ \ f_r(\cdots, z^{I}_i+a+b\tau, \cdots)
\\
&=
(-)^{-K_{II}(a+b)}
(-)^{\sum_J K_{IJ}N^{J}(a+b)}
 f_r(\cdots, z^{I}_i, \cdots) \times
\nonumber\\
&\ \ \ \ \ \ \
\e^{\imth K_{II} \tau \pi b^2  }
\e^{-\imth \sum_J K_{IJ} \{ \tau \pi b^2N^{J} + 2\pi  b 
[\sum_{j} (z^{I}_i - z^{J}_j) ]\} } 
\nonumber 
\end{align}
where $N^{I}$ is the total number electrons in the $I^\text{th}$ layer.

We know that $N^{I} \propto N_\phi$.
In large $N_\phi$ limit, the
$\e^{-\imth \tau \pi b^2N_\phi}$ term in \eqn{qpconMM}
must come from the $ \e^{-\imth \sum_J K_{IJ} \tau \pi b^2N^{J}}$ term in
\eqn{frabM1}. Thus $N^{I}$ and $N_\phi$ must satisfy
\begin{equation}
\label{NphiK}
 N_\phi=\sum_J K_{IJ} N^{J} .
\end{equation}
Now \eqn{frabM1} becomes
\begin{align}
\label{frabM}
&\ \ \ f_r(\cdots, z^{I}_i+a+b\tau, \cdots)
\\
&=
(-)^{-K_{II}(a+b)}
(-)^{\sum_J K_{IJ}N^{J}(a+b)}
 f_r(\cdots, z^{I}_i, \cdots) \times
\nonumber\\
&\ \ \ 
\e^{\imth K_{II} \tau \pi b^2  }
\e^{-\imth \{ \tau \pi b^2(\sum_J K_{IJ}N^{J}) + 2\pi  b 
[\sum_J K_{IJ} \sum_{j} (z^{I}_i - z^{J}_j) ]\} } 
\nonumber\\
&=f_r(\cdots, z_i,\cdots) 
\e^{-\imth \tau \pi b^2N_\phi -\imth 2\pi b N_\phi z^{I}_i}\times
\nonumber\\
&
(-)^{-K_{II}(a+b)}
(-)^{\sum_J K_{IJ}N^{J}(a+b)}
\e^{\imth K_{II} \tau \pi b^2 + \imth 2\pi  b\sum_J K_{IJ} Z^{J}}.
\nonumber 
\end{align}
We see that $f(\{z^{I}_i\})$ will satisfy the quasi periodic condition 
\eq{qpconMM} if $f_c(\{Z^{I}\})$ satisfies
\begin{align*}
&\ \ \
f_c(\cdots,Z^{I}+a+b\tau,\cdots) =
f_c(\cdots, Z^{I}, \cdots) \times
\\
&\ \ \ \  
(-)^{(a+b)(K_{II}-N_\phi)} 
\e^{-\imth K_{II} \tau \pi b^2 - \imth 2\pi  b\sum_J K_{IJ} Z^{J}}
,
\end{align*}
The above can be written in a vector form:
\begin{align}
\label{qpconfcMM1}
&\ \ \
f_c(\v Z+\v a+\v b\tau) 
\\
&=
f_c(\v Z^{I})  
(-)^{(\v \ka+N_\phi \v I)^T (\v a+\v b)} 
\e^{-\imth  \tau \pi \v b^T K \v b - \imth 2\pi  \v b^T K \v Z}
,
\nonumber 
\end{align}
where
\begin{equation}
\label{kaKII}
 \v\ka^T = (K_{11},K_{22},\cdots,K_{\ka\ka}) ,
\end{equation}
and $\v I^T=(1,\cdots,1)$.
From now on, we will assume 
\begin{align}
\label{NphiE}
N_\phi= \text{even}, 
\end{align}
and the above becomes
\begin{align}
\label{qpconfcMM}
&\ \ \ f_c(\v Z+\v a+\v b\tau)
\nonumber\\
&=
f_c(\v Z^{I})  
(-)^{\v\ka^T (\v a+\v b)} 
\e^{-\imth  \tau \pi \v b^T K \v b - \imth 2\pi  \v b^T K \v Z}
.
\end{align}


To find the center-of-mass wave function that satisfy the above quasi-periodic
condition, let us consider the following functions labeled by $\v \al,\v\eta$:
\begin{align}
\label{faletadef}
&\ \ \
f^{(\v\al,\v\eta)}(\v Z|\tau)
\\
& =
\sum_{\v n\in Z^\ka} \e^{ 
\imth\pi (\v n + \v\al+\v \eta)^TK\tau (\v n + \v\al+\v \eta)
+ \imth 2  \pi (\v n +\v\al+\v \eta)^T K \cdot(\v Z -\v \eta)},
\nonumber 
\end{align}
where the vector $\v \al$ satisfies
\begin{equation}
\label{Kal}
K\v\al= \text{integer vector} .
\end{equation}
We find that, for integral vectors $\v a$ and $\v b$,
\begin{align}
\label{faletaab1}
&\ \ \  f^{(\v\al,\v\eta)}(\v Z+\v a+\v b \tau|\tau)
\\
&=\sum_{\v n\in Z^\ka} 
\e^{ \imth\pi (\v n + \v\al+\v \eta)^TK\tau (\v n + \v\al +\v \eta)
+ \imth 2  \pi (\v n +\v\al+\v\eta)^T K \cdot(\v Z+\v a+\v b\tau -\v \eta) },
\nonumber\\
&=
\e^{\imth 2\pi\v\eta^T K \v a}
\e^{-\imth\pi \v b^TK\tau \v b}
\times
\nonumber\\
&
\sum_{\v n\in Z^\ka} 
\e^{ \imth\pi (\v n +\v b+ \v\al+\v\eta)^TK\tau (\v n + \v b+\v\al+\v\eta)
+ \imth 2  \pi (\v n +\v\al+\v\eta)^T K \cdot(\v Z-\v\eta)}
\nonumber\\
&=
\e^{\imth 2\pi\v\eta^T K \v a+\imth 2\pi\v\eta^TK \v b}
\e^{-\imth\pi \v b^TK\tau \v b -\imth 2  \pi \v b^T K \cdot\v Z }
f^{(\v\al,\v\eta)}(\v Z|\tau) .
\nonumber 
\end{align}

For bosonic electrons, $K_{II}=$ even
and $(-)^{\v\ka^T (\v a+\v b)}=1$.
Thus, the center-of-mass wave function $f_c(\v Z)$ is given by
\begin{equation}
\label{falcB}
 f^{(\v\al)}_c(\v Z)=\eta^{-\ka}(\tau)f^{(\v\al,\v 0)}(\v Z)
\end{equation}
For fermionic electrons, $K_{II}=$ odd
and $(-)^{\v\ka^T (\v a+\v b)}\neq 1$.
The center-of-mass wave function $f_c(\v Z)$ will be
\begin{equation}
\label{falcF}
 f^{(\v\al)}_c(\v Z)=\eta^{-\ka}(\tau)f^{(\v\al,K^{-1}\v\ka/2)}(\v Z)
\end{equation}
Note that here, we include an extra non-zero normalization factor
\begin{align}
 \eta(\tau)=q^{1/24}\prod_{n=1}^{\infty}(1-q^n) \Big|_{q=\e^{\imth 2\pi\tau}}
\end{align}
which is a holomorphic function of $\tau$ and satisfies
\begin{align}
\label{etatrans}
 \eta(\tau+1)=\e^{\imth\pi/12} \eta(\tau),\ \ \ \
 \eta(-1/\tau)=\sqrt{-\imth\tau}\eta(\tau).
\end{align}
Each center-of-mass wave function $f^{(\v\al)}_c(\v Z)$ 
gives rise to a degenerate
ground state $|\v\al;\tau\>$ (where $\tau$ describes the shape
of the torus).  
The explicit wave wave for the $|\v\al;\tau\>$ state is
\begin{align}
\label{PsiKal}
 \Psi_K^{\v\al}(\{z^{I}_i\}) &= 
f^{(\v\al)}(\{z^{I}_i\}) \e^{-\frac{B}{2}\sum_{i,I} [y^{I}_i]^2}
\\
f^{(\v\al)}(\{z^{I}_i\})&= f^{(\v\al)}_c(\{Z^{I}\}) f_r(\{z^{I}_i\})
\nonumber\\
f_r(\{z^{I}_i\}) &=
\Big\{
\prod_{I<J}
\prod_{i,j} \eta^{-K_{IJ}}(\tau)\th^{K_{IJ}}_{11}(z^{I}_i-z^{J}_j|\tau)
\Big\}
\times
\nonumber\\
&\ \ \ \ \
\Big\{
\prod_{I}
\prod_{i<j} \eta^{-K_{II}}(\tau)\th^{K_{II}}_{11}(z^{I}_i-z^{I}_j|\tau)
\Big\}
\nonumber 
\end{align}
From the definition \eq{faletadef}, we see that
\begin{equation*}
 f^{(\v\al+\v a,\v\eta)}(\v
Z|\tau)=f^{(\v\al,\v\eta)}(\v
Z|\tau),\ \ \ \ \ \ \ \ \v a=\text{ integer vector} .
\end{equation*}
Thus the distinct functions $f^{(\v\al,\v\eta)}(\v Z|\tau)$ are labeled by
$\v\al$ in the unit $\ka$ dimensional cube which will be denoted as UC.  The
number of different $\v\al$ in the unit $\ka$ dimensional cube, $\v\al\in UC$,
that satisfy \eqn{Kal} is $|\det(K)|$.  Thus \emph{an Abelian FQH labeled by
$K$ has $|\det(K)|$-fold degenerate ground states}.

Note that
\begin{align*}
 \sum_{i,I} (y_i^{I})^2
=\frac{1}{N} \Big[(\sum_{i,I} y_i^{I})^2
+ \frac 12 \sum_{i,I;j,J} (y_i^{I}-y_j^{J})^2 \Big]
\end{align*}
where $N=\sum_I N^{I}$ is the total number of electrons.  So the center of
mass motion described $\v Z$ and the relative motion described by $ z_i^{I}-
z_j^{J}$ totally factorize in the ground state wave function.

Also note that the wave function for the center of mass motion $
\Psi_{K,c}^{\v\al}(\v Z)  = f^{(\v\al)}_c(\{Z^{I}\})
\e^{-\frac{B}{2N}(\sum_{i,I} y^{I}_i)^2} $ form an orthogonal basis
\begin{align}
\label{PsiKcdel}
 \<\Psi_{K,c}^{\v\al}|\Psi_{K,c}^{\v\bt}\>\propto \del_{\v\al\v\bt}.
\end{align}
This can be derived bu considering generalized translation operators which
translate electrons in one or several layers.  Because $\Psi_{K,c}^{\v\al}(\v
Z)$ with different $\v\al$ are link by generalized translation operators which
are unitary operators, this allows us to obtain \eq{PsiKcdel}.  A discussion
of the translations that translate all electrons together is given in the next
section.  Since the center of mass motion  and the relative motion
separate, the many-body wave functions $\Psi_{K}^{\v\al}(\{z_i^{I} \})$ are
also orthogonal:
\begin{align}
\label{Psinorm}
 \<\Psi_{K}^{\v\al}|\Psi_{K}^{\v\bt}\>\propto \del_{\v\al\v\bt}.
\end{align}
We can always choose a normalization so that
$\<\Psi_{K}^{\v\al}|\Psi_{K}^{\v\bt}\>= \del_{\v\al\v\bt}$.  But in this case,
$|\Psi_{K}^{\v\bt}\>$ will not be a holomorphic function of $\tau$.  In this
paper, we will choose a ``holomorphic normalization'' such that
$|\Psi_{K}^{\v\bt}\>$ are holomorphic function of $\tau$ and satisfy
\eqn{Psinorm}.

\subsection{Translation symmetry}

The magnetic translation that acts on the holomorphic part $f(\{z^{I}_i\})$
is given by
\begin{equation}
\label{tTdNK}
\t T^{(N)}_{\v d}=\prod_{i,I} \e^{\imth d_y B (z^{I}_i+ \frac{d_x+\imth d_y}{2})}
\e^{\v d \cdot \v \prt^{I}_i} .
\end{equation}
Again, on torus, the allowed magnetic translations are generated by
$\t T^{(N)}_1=\t T^{(N)}_{(N_\phi^{-1},0)}$ and
$\t T^{(N)}_2=\t T^{(N)}_{(\tau_x N_\phi^{-1},\tau_y N_\phi^{-1})}$.
$\t T^{(N)}_1$ and
$\t T^{(N)}_2$ generate
the Heisenberg algebra \eq{HalgN}
where $N$ is the total number of electrons on all the layers.
Again $\t T^{(N)}_{\v d}$ acts only on the center-of-mass wave function
$f_c(\v Z)$.
From \eqn{faletatrans}
we find, for bosonic electrons,
\begin{align}
\label{T12falcB}
&\ \ \ \
\t T^{(N)}_1 f^{(\v\al)}_c(\v Z|\tau)=
f^{(\v\al)}_c(\v Z+\frac{\v N}{N_\phi}|\tau)
\nonumber\\
&=\e^{\imth 2\pi \v I^T\v\al}f^{(\v\al)}_c(\v Z|\tau) ,
\\
&\ \ \ \
\t T^{(N)}_2
f^{(\v\al)}_c(\v Z|\tau)=
\e^{ \imth 2 \pi \v Z^T \v I +\imth \frac{\pi \tau N}{N_\phi}}
f^{(\v\al)}_c(\v Z+\frac{\v N}{N_\phi}\tau|\tau)
\nonumber\\
&=f^{(\v\al+K^{-1}\v I)}_c(\v Z|\tau) .
\end{align}
and for fermionic electrons,
\begin{align}
\label{T12falcF}
& \ \ \
\t T^{(N)}_1 f^{(\v\al)}_c(\v Z|\tau)=
f^{(\v\al)}_c(\v Z+\frac{\v N}{N_\phi}|\tau)
\nonumber\\
&=\e^{\imth 2\pi (\v\al+\v\ka/2)^T\v I}f^{(\v\al)}_c(\v Z|\tau) 
\nonumber\\
&=(-)^{\v\ka^T\v I} \e^{\imth 2\pi \v I^T\v\al}f^{(\v\al)}_c(\v Z|\tau) ,
\\
&\ \ \ \
\t T^{(N)}_2
f^{(\v\al)}_c(\v Z|\tau)=
\e^{ \imth 2 \pi \v Z^T \v I +\imth \frac{\pi \tau N}{N_\phi}}
f^{(\v\al)}_c(\v Z+\frac{\v N}{N_\phi}\tau|\tau)
\nonumber\\
&=(-)^{\v\ka^T\v I} f^{(\v\al+K^{-1}\v I)}_c(\v Z|\tau) .
\end{align}
where we have used 
\begin{align}
\label{vN}
K^{-1}\v I &=\v N/N_\phi.
\nonumber\\
\v N^T&=(N_{1},\cdots,N_{\ka}),
\\
N&=\sum_I N_{I} = \text{ the total number of electrons}.
\nonumber 
\end{align}
Here $N_I$ is the number of $I^\text{th}$ electrons.

We know that $f^{(\v\al)}_c(\v Z) f_r(\{z_i^{I})$ for $\v\al\in UC$
corresponds to one of the $|\det(K)|$ degenerate ground states, $|\v\al;\tau\>$.
Those degenerate ground states form a representation of
the Heisenberg algebra \eq{HalgN}. We find that,
for bosonic electrons,
\begin{align}
\label{T12alB}
T^{(N)}_1 |\v\al;\tau\> &=\e^{\imth 2\pi \v q^T\v\al}|\v\al;\tau\>, 
\nonumber\\
T^{(N)}_2 |\v\al;\tau\> &=|\v\al+K^{-1}\v q;\tau\>, 
\end{align}
and for fermionic electrons,
\begin{align}
\label{T12alF}
T^{(N)}_1 |\v\al;\tau\> &=(-)^{\v\ka^T\v q} \e^{\imth 2\pi \v q^T\v\al}|\v\al;\tau\>, 
\nonumber\\ 
T^{(N)}_2 |\v\al;\tau\> &=(-)^{\v\ka^T\v q}|\v\al+K^{-1}\v q;\tau\>,
\end{align}
where we have used $\v q=\v I$  and assumed $N_\phi=$ even.

\section{
Modular transformations of generic Abelian FQH states 
}


As we have discussed, the ground state degeneracy on compact spaces contain a
lot of information about the topological order.  However, those information is
not enough to completely characterize the topological order; \ie FQH states
with really different topological orders may have the same ground state
degeneracy on any compact spaces. To obtain a complete quantitative theory for
topological order, the first thing we need is to have a complete
characterization of topological order; \ie to find a labeling scheme that can
label all distinct topological orders.

One way to find such a complete labeling scheme is to use the non-Abelian
geometric phase associated with the degenerate ground states.\cite{Wrig,KW9327} This can
be done by deforming the mass matrix.  So let us consider the following
one-electron Hamiltonian on a torus $(X^1,X^2)\sim (X^1+1,X^2)\sim (X^1,X^2+1)$
with a general mass matrix:
\begin{equation}
\label{HAXAY}
   H_{X} = - \frac{1}{2}\sum_{i,j=1,2}
\Big(\frac{\partial}{\partial X^i} - \imth A_i \Big) 
g_{ij} 
\Big(\frac{\partial}{\partial X^j} - \imth A_j \Big) ,
\end{equation}
where $g$ is the inverse-mass-matrix and
\begin{equation}
\label{AXAY}
(A_1,A_2)=(-2\pi N_\phi X^2, 0)
\end{equation}
gives rise to a uniform magnetic field with $N_\phi$ flux quanta going through
the torus.  Let $z=x+\imth y=X^1+X^2\tau$ where $\tau=\tau_x+\imth \tau_y$ is a
complex number.  We find that
\begin{equation*}
 x=X^1+\tau_x X^2,\ \ \ \ \ \ \ \ \ \ \ y=\tau_y X^2
\end{equation*}
and
\begin{align*}
\prt_{X^1}&=\prt_x, 
&
\prt_{X^2}&=\tau_x \prt_x +\tau_y\prt_y, 
\nonumber\\
\prt_x&=\prt_{X^1}
&
\prt_y&= -\frac{\tau_x}{\tau_y}\prt_{X^1} +\frac{1}{\tau_y} \prt_{X^1}
\end{align*}
Let us choose the inverse-mass-matrix $g_{ij}$ such that,
in terms of $(x,y)$, the Hamiltonian \eq{HAXAY} has a simple form
\begin{align*}
\label{HAxAy1}
&   H_X = - \frac{\tau_y}{2} \left [
(\frac{\partial}{\partial x} - i A_{x} )^2 +
(\frac{\partial}{\partial y} - i A_{y} )^2 \right ] 
\nonumber\\
     &= - \frac{\tau_y}{2} \left [
(\frac{\partial}{\partial X^1} - i A_{x} )^2 +
(
-\frac{\tau_x}{\tau_y}\frac{\partial}{\partial X^1} 
+\frac{1}{\tau_y}\frac{\partial}{\partial X^2} 
- i A_{y} )^2 
\right ] 
\nonumber\\
     &= - \frac{\tau_y}{2} \left [
(1+\frac{\tau_x^2}{\tau_y^2})(\frac{\partial}{\partial X^1} - i A_1 )^2 +
\frac{1}{\tau_y^2}(\frac{\partial}{\partial X^2} - i A_2 )^2 
\right . 
\nonumber\\
&\ \ \ \ \ \ \
-\frac{\tau_x}{\tau_y^2}
(\frac{\partial}{\partial X^1} - i A_1)
(\frac{\partial}{\partial X^2} - i A_2 )
\nonumber\\
&\ \ \ \ \ \ \
\left.
-\frac{\tau_x}{\tau_y^2}
(\frac{\partial}{\partial X^2} - i A_2)
(\frac{\partial}{\partial X^1} - i A_1 )
\right]
.
\end{align*}
We find that the inverse-mass-matrix must be
\begin{equation}
\label{mtau}
 g(\tau)=
\bpm
\tau_y+\frac{\tau_x^2}{\tau_y} & -\frac{\tau_x}{\tau_y}\\
-\frac{\tau_x}{\tau_y} & \frac{1}{\tau_y}
\epm .
\end{equation}
and $(A_x,A_y)$ must satisfy
\begin{align*}
A_x-\frac{\tau_x}{\tau_y} A_y 
& = (1+\frac{\tau_x^2}{\tau_y^2})A_1-\frac{\tau_x}{\tau_y^2}  A_2
\nonumber\\
\frac{1}{\tau_y} A_y 
& = -\frac{\tau_x}{\tau_y^2}A_1+\frac{1}{\tau_y^2}  A_2
\end{align*}
We find that
\begin{align*}
 A_x&= -2\pi N_\phi \frac{y}{\tau_y}  ,
&
 A_y&=\frac{\tau_x}{\tau_y^2}  2\pi N_\phi y  .
\end{align*}
Therefore
\begin{align}
\label{HAxAy2}
&\ \ \
\e^{-\imth\pi N_\phi y^2\tau_x/\tau_y^2 }
H_X
\e^{ \imth\pi N_\phi y^2\tau_x/\tau_y^2 }
\nonumber\\
&= - \frac{\tau_y}{2} \left [
(\frac{\partial}{\partial x} - i (-)\frac{2\pi y}{\tau_y} N_\phi )^2 +
(\frac{\partial}{\partial y} )^2 \right ] 
\end{align}
which is equal to the Hamiltonian \eq{HAxAy} (up to the
the scaling factor $\tau_y$).  We see that the Hamiltonian
\eq{HAXAY} on torus $ (X^1,X^2)\sim (X^1+1,X^2)\sim (X^1,X^2+1) $ with a
$\tau$-dependent inverse-mass-matrix \eq{mtau} is closely related to the
Hamiltonian \eq{HAxAy} on torus $z\sim z+1\sim z+\tau$ with a diagonal
mass matrix.

%
%
%
%

The ground state of \eqn{HAxAy} has a form $f(z)\e^{-\pi N_\phi y^2/\tau_y}$.
This allows us to find the ground state of \eqn{HAXAY}:
\begin{align}
\label{PsiX}
 \Phi(X^1,X^2)&=\Big[
f(X^1+\tau X^2) 
\e^{-\pi N_\phi y^2/\tau_y} \Big]
\e^{\imth\pi N_\phi y^2\tau_x/\tau_y^2}
\nonumber\\
&= f(X^1+\tau X^2) \e^{\imth\pi N_\phi \tau (X^2)^2}
\end{align}

\Eqn{PsiX} relates the one-electron wave function on torus $ (X^1,X^2)\sim
(X^1+1,X^2)\sim (X^1,X^2+1) $, $\Phi(X^1,X^2)$, with the one-electron wave
function on torus $z\sim z+1\sim z+\tau$, $f(z)\e^{-\pi N_\phi y^2/\tau_y}$.
Such a result can be generalized to the multilayer Abelian FQH state 
for many electrons.  From the Abelian FQH state on torus $z\sim z+1\sim z+\tau$
\eq{PsiKth}, we find that the Abelian FQH state on torus $ (X^1,X^2)\sim
(X^1+1,X^2)\sim (X^1,X^2+1) $ is described by (see \eqn{PsiKth}, \eqn{falcB},
and \eqn{falcF})
\begin{align}
\label{PsiKX}
&\ \ \
 \Phi^{\v\al}_K(\{ X^{1,I}_{i},X_{i}^{2,I} \};\tau) 
\nonumber\\
&= 
f^{\v\al}(\{ X^{1,I}_{i}+\tau X_{i}^{2,I} \}) 
\e^{\imth\pi N_\phi \tau \sum_{i,I} [X^{2,I}_{i}]^2},
\nonumber\\
&
f^{\v\al}(\{ z^{I}_{i} \})=
f_c^{\v\al}(\{ Z^{I} \}) f_r(\{ z^{I}_{i} \})
\end{align}
We see that the $\tau$-dependent inverse-mass-matrix \eq{mtau} leads to
a $\tau$-dependent ground state wave-functions \eq{PsiKX}
for the degenerate ground states.

The wave functions \eq{PsiKX} form a basis of the degenerate ground states. As
we change the mass matrix or $\tau$, we obtain a family of basis parametrized
by $\tau$.  The family of basis can give rise to non-Abelian geometric
phase\cite{WZ8411} which contain a lot information on topological order in
the FQH state.  In the following, we will discuss such a non-Abelian  geometric
phase in a general setting. We will use $\ga$ to label the degenerate ground
states.

To find the non-Abelian geometric phase, let us first define parallel
transportation of a basis.  Consider a path $\tau(s)$ that deform the
inverse-mass-matrix $g(\tau_1)$ to $g(\tau_2)$: $\tau_1=\tau(0)$  and
$\tau_2=\tau(1)$.  Assume that for each inverse-mass-matrix $g[\tau(s)]$, the
many-electron Hamiltonian on torus $ (X^1,X^2)\sim (X^1+1,X^2)\sim (X^1,X^2+1)
$ has $\ka$-fold degenerate ground states $|\ga;s\>$, $\ga=1,\cdots,\ka$, and
a finite energy gap for excitations above the ground states.  We can always
choose a basis $|\ga;s\>$ for the ground states such that the basis for
different $s$ satisfy 
\begin{equation*}
\<\ga';s|\frac{\dd}{\dd s} |\ga;s\> =0.  
\end{equation*} 
Such a choice of basis $|\ga;s\>$ defines a parallel transportation from the
bases for inverse-mass-matrix $g(\tau_1)$ to that for inverse-mass-matrix
$g(\tau_2)$ along the path $\tau(s)$.


In general, the parallel transportation is path dependent.  If we choose
another path $\tau'(s)$ that connect $\tau_1$ and $\tau_2$, the parallel
transportation of the same basis for inverse-mass-matrix  $g(\tau_1)$,
$|\ga;s=0\>=|\ga;s=0\>'$, may result in a different basis for
inverse-mass-matrix $g(\tau_2)$, $|\ga;s=1\>\neq |\ga;s=1\>'$.  The
different basis are related by a unitary transformation. Such a path dependent
unitary transformation is the non-Abelian geometric phase.\cite{WZ8411}

However, for the degenerate ground states of a topologically ordered state
(including a FQH state), the  parallel transportation has a special property
that, up to a total phase, it is path independent (in the thermal dynamical
limit).  The parallel transportations along different paths
connecting $\tau_1$ and $\tau_2$ will change a basis for inverse-mass-matrix
$g(\tau_1)$ to the same basis for inverse-mass-matrix $\tau(\tau_2)$
up to an overall phase: $|\ga;s=1\>=\e^{\imth\phi} |\ga;s=1\>'$.
In particular, if we deform a inverse-mass-matrix through a loop into itself
(\ie $\tau(0)=\tau(1)$), the basis $|\ga;0\>$ will parallel transport into
$|\ga;1\>=\e^{\imth\vphi}|\ga;0\>$.  Thus, non-Abelian geometric phases for the
degenerate states of a topologically ordered state are only path-dependent
Abelian phases $\e^{\imth\vphi}$ which do not contain much information of
topological order.

\begin{figure*}
\centerline{
\includegraphics[width=5.0in]{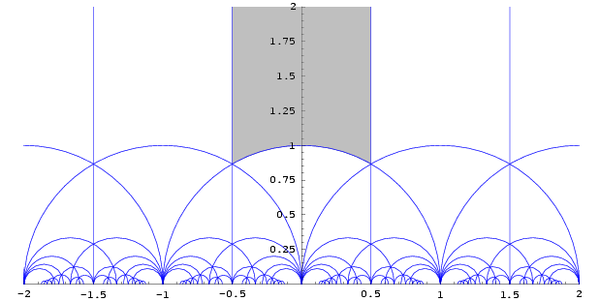}
}
\caption{(Color online)
Modular transformations generate by $\tau \to \tau+1$ and $\tau \to -1/\tau$
map $\tau$'s in one domain to $\tau$'s in another domain.  $\tau$'s in one
domain label distinct systems.  $\tau$'s in different domains related by a
modular transformation describe the same system (up to a coordinate
transformation).  Each domain has a topology of a sphere minus a point.
(The figure comes from Wikipedia.)
}
\label{MGDomain}
\end{figure*}

However, there is a class of special paths which give rise to non-trivial
non-Abelian geometric phases.  First we note that the torus $(X^1,X^2)\sim
(X^1+1,X^2) \sim (X^1,X^2+1)$ can be parameterized by another set of
coordinates
\begin{align}
\label{XpX}
& 
\bpm X^{\prime 1}\\ X^{\prime 2}\\ \epm
=
\bpm
d&-b\\
-c&a\\
\epm
 \bpm X^1\\ X^2\\ \epm,
\ \ 
\bpm X^1\\ X^2\\ \epm
=
\bpm
a&b\\
c&d\\
\epm
 \bpm X^{\prime 1}\\ X^{\prime 2}\\ \epm,
\end{align}
where $a,b,c,d \in Z$,  $ad-bc=1 $.
The above can be rewritten in vector form
\begin{align}
\label{XpXV}
& 
\v X = M\v X',
\ \ \ \ 
M=\bpm
a&b\\
c&d\\
\epm
\in SL(2,Z) .
\end{align}
$(X^{\prime 1},X^{\prime 2})$ has the same periodic condition $(X^{\prime 1},X^{\prime
2})\sim (X^{\prime
1}+1,X^{\prime 2})
\sim (X^{\prime 1},X^{\prime 2}+1)$ as that for $(X^1,X^2)$.  We note that 
\begin{equation*}
  \bpm \prt_{X^1}\\ \prt_{X^2}\\ \epm
=
\bpm
d&-c\\
-b&a\\
\epm
 \bpm \prt_{X^{\prime 1}}\\ \prt_{X^{\prime 2}}\\ \epm 
,\ \ \ \ 
  \bpm \prt_{X^{\prime 1}}\\ \prt_{X^{\prime 2}}\\ \epm
=
\bpm
a&c\\
b&d\\
\epm
 \bpm \prt_{X^1}\\ \prt_{X^2}\\ \epm
.
\end{equation*}
The inverse-mass-matrix in the $(X^1,X^2)$ coordinate, $g(\tau)$,
is changed to
\begin{equation*}
 g'= 
\bpm
d&-b\\
-c&a\\
\epm
g(\tau)
\bpm
d&-c\\
-b&a\\
\epm
\end{equation*}
in the $(X^{\prime 1},X^{\prime 2})$ coordinate.
From \eqn{mtau}, we find that
\begin{equation}
 g'=
\bpm
\tau_y'+\frac{{\tau_x'}^2}{\tau_y'} & -\frac{\tau_x'}{\tau_y'}\\
-\frac{\tau_x'}{\tau_y'} & \frac{1}{\tau_y'}
\epm = g(\tau'),
\end{equation}
with
\begin{equation}
\label{tauptau}
 \tau'=\frac{b+d\tau}{a+c\tau} .
\end{equation}
The above transformation $\tau\to \tau'$ is the modular transformation.  We
see that if $\tau$ and $\tau'$ are related by the modular transformation, then
two inverse-mass-matrices $g(\tau)$ and $g(\tau')$ will actually describe the
same system (up to a coordinate transformation). (See Fig. \ref{MGDomain})

Let us assume that the path $\tau(s)$ connects two $\tau$'s,
$\tau(0)$ and $\tau(1)$, related by a
modular transformation $M=\bpm a&b\\ c&d \epm$:
$\tau(1)=\frac{b+d\tau(0)}{a+c\tau(0)}$ (see Fig. \ref{MGDomain}).  We will denote $\tau(0)=\tau$ and
$\tau(1)=\tau'$.  The parallel transportation of the basis $|\ga;\tau\>$ for
inverse-mass-matrix $g[\tau(0)]$ gives us a basis $|\ga;\tau'\>$ for inverse-mass-matrix
$g[\tau(1)]$.  Since $\tau=\tau(0)$ and $\tau'=\tau(1)$ are related by a
modular transformation, $g[\tau(0)]$ and $g[\tau(1)]$ actually describe the
same system. The two basis $|\ga;\tau\>$ and $|\ga;\tau'\>$ are actually two
basis of same space of the degenerate ground states. Thus there is a unitary
matrix that relate the two basis
\begin{align}
\label{alUbt}
 |\ga;\tau'\> &=\hat U(M) |\ga;\tau\>
\\
U_{\ga\ga'}(M)
&
=\<\ga;\tau|\ga';\tau'\>
=\<\ga;\tau|\hat U(M)|\ga';\tau\>
.
\nonumber
\end{align}
Such a unitary matrix is the non-Abelian geometric phase for the path $\tau(s)$.

Let us consider a modular transformation generated by $M'=\bpm a'&b'\\ c'&d'
\epm$: $\tau' \to \tau''=\frac{b'+d'\tau'}{a'+c'\tau'}$.
We see that
\begin{align}
  \tau''=\frac{b'+d'\tau'}{a'+c'\tau'}
=\frac{(a b'+b d')+(c b' + d d') \tau}{
(a a' +b c')+(c a'+d c')\tau}
\end{align}
and the transformation $\tau\to \tau''$ is generated by
$M M'$. From $ |\ga;\tau''\> = \hat U(M') |\ga;\tau'\> 
= \hat U(M') \hat U(M) |\ga;\tau\> $, we find that
\begin{align}
\label{MMpUUp}
&\ \ \ U_{\ga\ga''}(MM')=\<\ga;\tau|\ga'';\tau''\>
\\
&=\sum_{\ga'} \<\ga;\tau|\ga';\tau'\> \<\ga';\tau'|\ga'';\tau''\>
=U_{\ga\ga'}(M) U_{\ga'\ga''}(M') 
\nonumber 
\end{align}

Except for its overall phase (which is path dependent), the unitary matrix $U$
is a function of the modular transformation $M$.  From \eqn{MMpUUp}, we see
that the unitary matrix $U$ form a projective representation of the modular
transformation.  The projective representation of the modular transformation
contains a lot of information (may even all information) of the underlying
topological order.

Let us examine $\<\ga';\tau'|\ga;\tau\>$ in \eqn{alUbt} more carefully.  Let
$\Phi_{\ga}[\{\v X_i^{I}\}|\tau]$ be ground state wave functions for
inverse-mass-matrix $g[\tau]$, and $\Phi_{\ga}[\{\v X_i\}|\tau']$ be ground
state wave functions for inverse-mass-matrix $g[\tau']$.  Here $\v
X_i=(X^{1,I}_i,X^{2,I}_i)$ are the coordinates of the $i^\text{th}$
electron in the $I^\text{th}$ component.  Since $\tau$ and $\tau'$ are related
by a modular transformation, $\Phi_{\ga}[\{\v X_i^{I}\}|\tau]$ and $\Phi_{\ga}[\{\v
X_i^{I}\}|\tau']$ are ground state wave function of the same system.  However, we
cannot directly compare $\Phi_{\ga}[\{\v X_i^{I}\}|\tau]$ and $\Phi_{\ga}[\{\v
X_i^{I}\}|\tau']$ and calculate the inner product between the two wave functions as 
\begin{equation*}
\int \prod_{i,I}\dd^2 \v
X_i^{I}\; \Big[\Phi_{\ga}[\{\v X_i^{I}\}|\tau(0)]\Big]^*\Phi_{\ga'}[\{\v
X_i^{I}\}|\tau'].  
\end{equation*}
The wave function $\Phi_{\ga}[\{\v X_i^{I}\}|\tau']$ for inverse-mass-matrix
$g[\tau']$ can be viewed as the ground state wave function for
inverse-mass-matrix $g[\tau]$ only after a coordinate transformation.  Let us
rename $\v X$ to $\v X'$ and rewrite $\Phi_{\ga}[\{\v X_i^{I}\}|\tau']$ as
$\Phi_{\ga}(\{\v X^{\prime,I}_i\}|\tau')$.  Since the coordinate
transformation \eq{XpXV} change $\tau$ to $\tau'$, we see that we should really
compare $\Phi_{\ga}(\{\v X^{\prime,I}_i\}|\tau')=\Phi_{\ga}(\{M^{-1}\v
X_i^{I}\}|\tau')$ with $\Phi_{\ga}(\{\v X_i^{I}\}|\tau)$. 
But even $\Phi_{\ga}(\{\v X_i^{I}\}|\tau)$ and
$\Phi_{\ga}(\{M^{-1}\v X_i^{I}\}|\tau')$ cannot be directly compared.  This
is because the coordinate transformation \eq{XpX} changes the gauge potential
\eq{AXAY} to another gauge equivalent form.  We need to perform a $U(1)$ gauge
transformation $U_G(M)$ to transform the changed gauge potential back to its
original form \eqn{AXAY}. So only
$\Phi_{\ga}[\{\v X_i^{I}\}|\tau]$ and
$U_G\Phi_{\ga}[\{M^{-1}\v X_i^{I}\}|\tau']$ can be directly compared.
Therefore, we have
\begin{align}
\label{PsiUPsi}
&\ \ \ U_{\ga\ga'}(M)
\\
&=\int_0^1 \prod_{i,I} \dd^2 \v X_i^{I}\; 
\Phi^*_{\ga}(\{\v X_i^{I}\}|\tau) 
 U_G(M)\Phi_{\ga'}(\{M^{-1}\v X_i^{I}\}|\tau')
\nonumber 
\end{align}
which is \eqn{alUbt} in wave function form.
Note that $\tau'=\frac{M_{12}+M_{22}\tau}{M_{11}+M_{21}\tau}$.

Let us calculate the gauge transformation $U_G(M)$.
We note that
$\Phi_{\ga}(\{\v X^{\prime,I}_i\}|\tau')$ is the ground state of
\begin{equation*}
 H'=-\sum_{k,I}
\frac{1}{2}\sum_{i,j=1,2}
\Big(\frac{\partial}{\partial X^{\prime i,I}_k} - \imth A'_i \Big) 
g'_{ij} 
\Big(\frac{\partial}{\partial X^{\prime j,I}_k} - \imth A'_j \Big) 
\end{equation*}
where $k=1,\cdots,N$ labels the different electrons.
In terms of $X^i$ (see \eqn{XpX}), $H'$ has a form  
\begin{equation*}
 H'=-\sum_{k,I}
\frac{1}{2}\sum_{i,j=1,2}
\Big(\frac{\partial}{\partial X^{i,I}_k} - \imth \t A_i \Big) 
g_{ij} 
\Big(\frac{\partial}{\partial X^{j,I}_k} - \imth \t A_j \Big) 
\end{equation*}
where
\begin{equation*}
  \bpm \t A_1\\ \t A_2\\ \epm
=
\bpm
d&-c\\
-b&a\\
\epm
 \bpm A'_1\\ A'_2\\ \epm 
,\ \ \ \ 
  \bpm A'_1 \\ A'_2\\ \epm
=
\bpm
a&c\\
b&d\\
\epm
 \bpm \t A_1 \\ \t A_2\\ \epm
.
\end{equation*}
Since
$(A'_1,A'_2)=(-2\pi N_\phi X^{\prime 2},0)$,
We find that
\begin{align*}
&\ \ \ \ (\t A_1,\t A_2)=
( -2\pi N_\phi X^{\prime 2} d,\ 2\pi N_\phi X^{\prime 2} b)
\nonumber\\
&=
( -2\pi N_\phi (-c X^1+ aX^2) d,\ 2\pi N_\phi (-c X^1+ aX^2) b)
\end{align*}
$U_G(M)$ will change $H'$ to $H$:
\begin{align*}
&\ \ \ \ U_G(M) H' U_G^\dag(M)=H
\nonumber\\
&=
-\sum_{k,I}
\frac{1}{2}\sum_{i,j=1,2}
\Big(\frac{\partial}{\partial X^{i,I}_k} - \imth A_i \Big) 
g_{ij} 
\Big(\frac{\partial}{\partial X^{j,I}_k} - \imth A_j \Big) 
\end{align*}
with $(A_1,A_2)=(-2\pi N_\phi X^2,0)$.
We find that
\begin{align}
\label{UGuG}
U_G(M;\{ \v X_i \}) &=\prod_{k,I} u_G(M;\v X_k^{I})  ,
\nonumber\\
u_G(M;\v X)&=\e^{\imth  2\pi N_\phi [
bc X^1X^2
-\frac{cd}{2} (X^1)^2
-\frac{ab}{2} (X^2)^2
] }  .
\end{align}
Note that at $\v X_i=0$, the fixed point of coordinate transformation
$\v X'_i=M^{-1}\v X_i$,
$u_G(M;\v X)|_{\v X=0} = 1$.

Eqn. \eq{PsiUPsi} can also be rewritten as a transformation
on the wave function $\Phi_{\ga}(\{\v X_i^{I}\}|\tau)$:
\begin{align}
\label{UPsi}
\hat U(M)\Phi_{\ga}(\{\v X_i^{I}\}|\tau)
&=
U_G(M)\Phi_{\ga}(\{\v X^{\prime,I}_i\}|\tau')
\nonumber\\
&=
U_G(M)\Phi_{\ga}(\{M^{-1}\v X_i^{I}\}|\tau')
\nonumber\\
&=
\Phi_{\ga'}(\{\v X_i^{I}\}|\tau)   U_{\ga'\ga}(M) .
\end{align}
where $\tau'=\frac{M_{12}+M_{22}\tau}{M_{11}+M_{21}\tau}$.  
%
Let 
\begin{equation*}
M_T=\bpm 1&1\\ 0&1 \epm,\ \ \ \ \ M_S=\bpm 0&-1\\ 1&0 \epm .  
\end{equation*}
Since $\Phi_{\ga}(\{\v X_i^{I}\}|\tau) $ has a form
\begin{align*}
 \Phi_{\ga}(\{\v X_i^{I}\}|\tau)
=f_{\ga}(\{X^{1,I}_i+\tau X^{2,I}_i \}|\tau) 
e^{\imth \pi \tau N_\phi \sum_i (X^{2,I}_i)^2}
\end{align*}
where $f_{\ga}(\{z_i^{I} \}|\tau)$ is a holomorphic function, we can express the
action of $\hat U(M_T)$ and $\hat U(M_S)$ on $\Phi_{\ga}(\{\v X_i^{I}\}|\tau) $ as 
\begin{widetext}
\begin{align*}
&
\hat U(M_T)\Phi_{\ga}(\{\v X_i^{I}\}|\tau)
=
U_G(M_T)\Phi_{\ga}(\{X^{1,I}_i-X^{2,I}_i,X^{2,I}_i\}|\tau+1)
\nonumber\\
&=
U_G(M_T)f_{\ga}(\{(X^{1,I}_i-X^{2,I}_i)+(\tau+1) X^{2,I}_i\}|\tau+1)
e^{\imth \pi (\tau+1) N_\phi \sum_{i,I} (X^{2,I}_i)^2} 
\nonumber\\
&=
\e^{-\imth  \pi N_\phi \sum_{i,I}(X^{2,I}_i)^2  }  
f_{\ga}(\{z_i^{I}\}|\tau+1)
e^{\imth \pi (\tau+1) N_\phi \sum_{i,I} (X^{2,I}_i)^2}
=
f_\ga ( \{ z_i^{I} \} |\tau + 1) 
e^{\imth \pi \tau N_\phi \sum_{i,I} (X^{2,I}_i)^2} ,
\end{align*}
and
\begin{align*}
& \hat U(M_S)\Phi_{\ga}(\{\v X_i^{I}\}|\tau)
=
U_G(M_S)\Phi_{\ga}(\{X_i^{2,I},-X_i^{1,I}\}|\frac{-1}{\tau})
=
U_G(M_T)f_{\ga}(\{X^{2,I}_i-(\frac{-1}{\tau}) X^{1,I}_i\}|\frac{-1}{\tau})
e^{\imth \pi (\frac{-1}{\tau}) N_\phi \sum_{i,I} (X^{1,I}_i)^2} 
\nonumber\\
&=
\e^{-2 \imth  \pi N_\phi \sum_{i,I} X^{1,I}_i X^{2,I}_i  }  
f_\ga ( \{ z_i/\tau\} |\frac{-1}{\tau})
 e^{-\imth \pi N_\phi \sum_{i,I} (X^{1,I}_i)^2/\tau} 
=
f_\ga ( \{ z_i^{I}/\tau\} |\frac{-1}{\tau})
\e^{-\imth \pi N_\phi \sum_{i,I} (z_i^{I})^2/\tau}
 e^{\imth \pi \tau N_\phi \sum_{i,I} (X^{2,I}_i)^2} ,
\end{align*}
\end{widetext}
where $z_i^{I}=X^{1,I}_i+\tau X^{2,I}_i$.
We see that the action of operator $\hat U(M)$ can be expressed as 
the action of another operator $\hat{\t U}(M)$
on $f_{\ga}(\{\v X_i^{I}\}|\tau) $, where
\begin{equation*}
\hat{\t U}(M)= 
 e^{-\imth \pi \tau N_\phi \sum_i (X^{2,I}_i)^2} 
\hat U(M)
 e^{\imth \pi \tau N_\phi \sum_i (X^{2,I}_i)^2} .
\end{equation*}
We find
\begin{align}
\label{tUMf}
\hat{\t U}(M_T)f_{\ga}(\{z_i^{I}\}|\tau)
&=
f_\ga ( \{ z_i^{I} \} |\tau + 1) 
\nonumber\\
&=
f_{\ga'} ( \{ z_i^{I} \} |\tau ) U_{\ga'\ga}(M_T)
,
\nonumber \\
\hat{\t U}(M_S)f_{\ga}(\{z_i^{I}\}|\tau)
&=
f_\ga ( \{ z_i^{I}/\tau\} |\frac{-1}{\tau})\e^{-\imth \pi N_\phi \sum_{i,I} (z_i^{I})^2/\tau} 
\nonumber\\
&
=
f_{\ga'} ( \{ z_i^{I} \} |\tau ) U_{\ga'\ga}(M_S)
.
\end{align}
We may use this result to calculate $U_{\ga\ga'}(M)$.

In the above, we have assumed that $\Phi_{\ga}(\tau)$ form an orthonormal basis
$\< \Phi_{\ga}(\tau)|\Phi_{\ga'}(\tau) \>=\del_{\ga\ga'}$.  In this case,
$U_{\ga\ga'}(M)$ is an unitary matrix. However, it is more convenient to 
choose ``chiral normalization'' such that $\<\Phi_{\ga}(\tau)|\Phi_{\ga'}(\tau)
\>\propto \del_{\ga\ga'}$ and that $\Phi_{\ga}(\tau)$ is a holomorphic function
of $\tau$. We will choose such a  ``chiral normalization'' in this paper.

To summarize, there are two kinds of deformation loops $\tau(s)$. If
$\tau(0)=\tau(1)$, the deformation loop is contactable [\ie we can deform the
loop to a point, or in other words we can continuously deform the function
$\tau(s)$ to a constant function $\tau(s)=\tau(0)=\tau(1)$].  For a
contactable loop, the associated non-Abelian geometric phase is actually a
$U(1)$ phase $U_{\ga\ga'}=\e^{\imth \vphi}\del_{\ga\ga'}$.  where $\vphi$ is
path dependent.  If $\tau(0)$ and $\tau(1)$ are related by a modular
transformation, the deformation loop is non-contactable.  Then the associated
non-Abelian geometric phase is non-trivial.  If two non-contactable loops can
be deformed into each other continuously, then the two loops only differ by a
contactable loop.  The associated non-Abelian geometric phases will only differ
by a an overall $U(1)$ phase.  Thus, up to an overall $U(1)$ phase, the
non-Abelian geometric phases $U_{\ga\ga'}$ of a topologically ordered state are
determined by the modular transformation
$\tau\to\tau'=\frac{b+d\tau}{a+c\tau}$.
We also show that we can use the
parallel transportation to defined a system of basis $\Phi_\ga(\{\v
X_i\}|\tau)$ for all inverse-mass-matrices labeled by $\tau$.

\subsection{Calculation for Abelian states}

For the Abelian FQH states, we have already introduced a system of basis for
the degenerate ground states for each inverse-mass-matrix $g(\tau)$:
$|\v\al;\tau\>=\Phi^{\v\al}_K(\{\v X^I_i\}|\tau)$
(see \eqn{falcB}, \eqn{falcF}, and \eqn{PsiKal}).  It appears that such a
system of basis happen to be a system of orthogonal basis that are related by
the parallel transformation (up to an overall phase), \ie the basis satisfy
\begin{equation}
\label{ptrans}
\<\v\bt;\tau|\frac{\dd}{\dd \tau} |\v\al;\tau\> \propto \del_{\v\al\v\bt},
\ \ \ \ 
\<\v\bt;\tau|\frac{\dd}{\dd \tau^*} |\v\al;\tau\> \propto \del_{\v\al\v\bt} .
\end{equation}
One way to see this result is to note that the wave function
$\Phi^{\v\al}_K(\{\v X^I_i\}|\tau)$ factorizes as a product of center-of-mass
wave function and relative-motion wave function (see \eqn{PsiKal}). So up to
the over all $U(1)$ phase, the non-Abelian geometric phases are determined by the
center-of-mass wave function only, because the  relative-motion wave function
does not depend on $\v\al$.  If we only look at the center-of-mass wave
functions $\Phi^{\v\al}_{K,c}(\v Z|\tau)$, one can show that they form a system
of orthogonal basis (parametrized by $\tau$) that are related by the parallel
transformation (up to an overall phase).

So if we choose the basis for the degenerate ground states as
$|\v\al;\tau\>=\Phi^{\v\al}_K(\{\v X^I_i\}|\tau)$ (see \eqn{falcB},
\eqn{falcF}, and \eqn{PsiKal}), then we can use \eqn{tUMf} to calculate the
non-Abelian geometric phases associated with the two generators $T: \tau \to
\tau+1$ and $S: \tau\to \-1/\tau$ of the modular transformations. 

Let us consider the bosonic electrons. From \eqn{PsiKal} and
\eqn{falcB}, we see that
\begin{align}
\label{bfal}
&  f^{(\v\al)}(\{z^{I}_i\}|\tau)= \eta^{-\ka}(\tau)f^{(\v\al,0)}(\{Z^{I}\}|\tau) \times
\\
&\ \ \ \ \ \ \ \ \ \ \
\Big\{
\prod_{I<J}
\prod_{i,j} \eta^{-K_{IJ}}(\tau)\th^{K_{IJ}}_{11}(z^{I}_i-z^{J}_j|\tau)
\Big\}\times
\nonumber\\
&\ \ \ \ \ \ \ \ \ \ \
\Big\{
\prod_{I}
\prod_{i<j} \eta^{-K_{II}}(\tau)\th^{K_{II}}_{11}(z^{I}_i-z^{I}_j|\tau)
\Big\}
\nonumber 
\end{align}
Using \eqn{falBtau1}, \eqn{taup1}, and \eqn{etatrans},
we obtain
\begin{align}
&  f^{(\v\al)}(\{z^{I}_i\}|\tau+1)
=
\e^{\frac{1}{6}\imth\pi [\sum_{I<J} K_{IJ}N^I N^J +
\sum_I \frac{K_{II}}{2} N_I(N_I-1)]} \times
\nonumber\\
&\ \ \ \ \ \ \ \ \ \ \ \ \ \ \ \ \ \ \ \ \ \  \ \ \ \ \ \ \
\e^{\imth \pi (\v\al^TK\v\al-\frac{1}{12}\ka)}
  f^{(\v\al)}(\{z^{I}_i\}|\tau)
\nonumber\\
&=
\e^{\frac{1}{12}\imth\pi  (\v N^TK \v N-\v \ka^T\v N )}
\e^{\imth \pi (\v\al^TK\v\al-\frac{1}{12}\ka)}
  f^{(\v\al)}(\{z^{I}_i\}|\tau)
\end{align}
where $\v N$ is a column vector with component $N_I$ and
$\v \ka$ is a column vector with component $K_{II}$ (see \eqn{vN}).
Therefore, from the definition \eqn{PsiKX},
we find that
\begin{align}
& \Phi^{\v\al}_K(\{\t X^{1,I}_i, \t X^{2,I}_i\};\tau+1)
=
\e^{\imth\pi N_\phi\sum_{i,I}(X_i^{2,I})^2}
\times
\\
&\ \ \ 
\e^{\frac{1}{12}\imth\pi  (\v N^TK \v N-\v \ka^T\v N )}
\e^{\imth \pi (\v\al^TK\v\al-\frac{1}{12}\ka)}
 \Phi^{\v\al}_K(\{X^{1,I}_i, X^{2,I}_i\};\tau)
\nonumber 
\end{align}
where
\begin{align}
 \t X^{1,I}_i+(\tau+1) \t X^{2,I}_i
=
 X^{1,I}_i+\tau X^{2,I}_i
\end{align}
or
\begin{align}
 X^{1,I}_i= \t X^{1,I}_i+ \t X^{2,I}_i,\ \ \ \
 X^{2,I}_i= \t X^{2,I}_i.
\end{align}

Using \eqn{falBtauinv}, \eqn{invtau}, and \eqn{etatrans},
we obtain that
\begin{align}
&  f^{(\v\al)}(\{z^{I}_i/\tau\}|-1/\tau)
=
(-)^{(\v N^TK \v N-\v \ka^T\v N )/2}\times
\nonumber\\
& \ \ \
\e^{\imth \frac{\pi}{\tau} [
\sum_{I<J;i,j} K_{IJ} (z^I_i-z^J_j)^2
+\sum_{I;i<j} K_{II}(z^I_i-z^I_j)^2]
 } \times
\\
&\ \ \
\e^{\imth\pi \v Z^TK \v Z/\tau }
\sum_{\v\bt\in UC}
\frac{\e^{ -\imth 2\pi \v\bt^T K\v\al }}
{ \sqrt{\det(K)} }
  f^{(\v\bt)}(\{z^{I}_i\}|\tau)
\nonumber 
\end{align}
Note that
\begin{align}
& \ \ \ \sum_{I<J;i,j} K_{IJ} (z^I_i-z^J_j)^2
+\sum_{I;i<j} K_{II} (z^I_i-z^I_j)^2
\nonumber\\
&=\frac12 \sum_{I,J;i,j} K_{IJ} (z^I_i-z^J_j)^2
=N_\phi \sum_{I,i} (z^I_i)^2 -\sum_{I,J;i,j} K_{IJ} z^I_i z^J_j
\nonumber \\
&
=N_\phi \sum_{I,i} (z^I_i)^2 -\v Z^T K \v Z,
\end{align}
where we have used $N_\phi=K_{IJ}N_J$.
Therefore, we have
\begin{align}
&\ \ \  f^{(\v\al)}(\{z^{I}_i/\tau\}|-1/\tau)
\e^{-\imth \frac{\pi}{\tau} N_\phi \sum_{I,i} (z^I_i)^2 }
\nonumber\\
&=
(-)^{(\v N^TK \v N-\v \ka^T\v N )/2}
\sum_{\v\bt\in UC}
\frac{\e^{ -\imth 2\pi \v\bt^T K\v\al }}
{ \sqrt{\det(K)} }
  f^{(\v\bt)}(\{z^{I}_i\}|\tau)
\nonumber 
\end{align}
{}From the definition \eqn{PsiKX},
we find that
\begin{align}
& \Phi^{\v\al}_K(\{\t X^{1,I}_i, \t X^{2,I}_i\};-1/\tau)
=
\e^{2\imth\pi N_\phi\sum_{i,I}X_i^{1,I}X_i^{2,I}}
\times
\\
&
(-)^{(\v N^TK \v N-\v \ka^T\v N)/2 }
\sum_{\v\bt\in UC}
\frac{\e^{ -\imth 2\pi \v\bt^T K\v\al }}
{ \sqrt{\det(K)} }
 \Phi^{\v\bt}_K(\{X^{1,I}_i, X^{2,I}_i\};\tau)
\nonumber 
\end{align}
where
\begin{align}
 \t X^{1,I}_i-\frac{1}{\tau} \t X^{2,I}_i
=
 (X^{1,I}_i+\tau X^{2,I}_i)/\tau
\end{align}
or
\begin{align}
 X^{1,I}_i= -\t X^{2,I}_i,\ \ \ \
 X^{2,I}_i= \t X^{1,I}_i.
\end{align}

So for bosonic electrons, the non-Abelian geometric phases
are described by the following generators of 
(projective) representation of modular group
\begin{align}
\label{bST}
 U_{\v\al\v\bt}(M_T)&=
\e^{\frac{\imth\pi}{12}  (\v N^TK \v N-\v \ka^T\v N )}
\e^{\imth \pi (\v\al^TK\v\al-\frac{1}{12}\ka)}\del_{\v\al\v\bt}
\nonumber\\
 U_{\v\al\v\bt}(M_S)&=
(-)^{(\v N^TK \v N-\v \ka^T\v N )/2}
\frac{\e^{ -\imth 2\pi \v\bt^T K\v\al }}
{ \sqrt{\det(K)} }
\end{align}

Now we consider the fermionic electrons. From \eqn{PsiKal} and \eqn{falcF}, 
we see that
\begin{align}
\label{ffal}
&  f^{(\v\al)}(\{z^{I}_i\}|\tau)= \eta^{-\ka}(\tau)f^{(\v\al,K^{-1}\v\ka/2)}(\{Z^{I}\}|\tau) \times
\\
&\ \ \ \ \
\Big\{
\prod_{I<J}
\prod_{i,j} \eta^{-K_{IJ}}(\tau)\th^{K_{IJ}}_{11}(z^{I}_i-z^{J}_j|\tau)
\Big\}\times
\nonumber\\
&\ \ \ \ \
\Big\{
\prod_{I}
\prod_{i<j} \eta^{-K_{II}}(\tau)\th^{K_{II}}_{11}(z^{I}_i-z^{I}_j|\tau)
\Big\}
\nonumber 
\end{align}
Using \eqn{falFtau1} and \eqn{taup1},
we obtain that
\begin{align}
&  f^{(\v\al)}(\{z^{I}_i\}|\tau+1)
=\e^{\frac{1}{12}\imth\pi  (\v N^TK \v N-\v \ka^T\v N -\ka)}\times
\nonumber\\
&\ \ \ \
\e^{\imth \pi (\v\al+\frac12 K^{-1}\v\ka)^T K (\v\al+\frac12 K^{-1}\v\ka)}
  f^{(\v\al)}(\{z^{I}_i\}|\tau).
\end{align}
Using \eqn{falFtauinv} and \eqn{invtau}, we obtain 
\begin{align}
&\ \ \  f^{(\v\al)}(\{z^{I}_i/\tau\}|-1/\tau)
\e^{-\imth \frac{\pi}{\tau} N_\phi \sum_{I,i} (z^I_i)^2 }
\nonumber\\
&=
(-)^{(\v N^TK \v N-\v \ka^T\v N )/2}
\e^{ \frac 12 \imth \pi \v\ka^T K^{-1} \v\ka }\times
\nonumber\\
&\ \ \ \ \ 
\sum_{\v\bt\in UC} 
\frac{ \e^{ -\imth 2\pi ( \v\bt^T K\v\al +\v\bt^T \v\ka ) } } { \sqrt{\det(K)} }
  f^{(\v\bt)}(\{z^{I}_i\}|\tau)
\nonumber 
\end{align}
So for fermionic electrons, the non-Abelian geometric phases
are described by the following generators of 
(projective) representation of modular group
\begin{align}
\label{fST}
 U_{\v\al\v\bt}(M_T)&=
\e^{\frac{1}{12}\imth\pi  (\v N^TK \v N-\v \ka^T\v N -\ka)}\times
\nonumber\\
&\ \ \ \
\e^{\imth \pi (\v\al+\frac12 K^{-1}\v\ka)^TK(\v\al+\frac12 K^{-1}\v\ka)}\del_{\v\al\v\bt}
\nonumber\\
 U_{\v\al\v\bt}(M_S)&=
(-)^{(\v N^TK \v N-\v \ka^T\v N )/2} \times
\nonumber\\
&\ \ \ \
\e^{ \frac12 \imth \pi \v\ka^T K^{-1} \v\ka }
\frac{ \e^{ -\imth 2\pi ( \v\al^T K\v\bt +\v\bt^T \v\ka ) } } { \sqrt{\det(K)} }
\end{align}

\section{Summary}

Let us summarize the results obtained in this paper.  For an Abelian FQH state
$(K, \v q)$ on a torus, the ground state degeneracy is given by $|\det(K)|$. The
degenerate ground states $|\v\al\>$ can be labeled by the $\ka$ dimensional
vectors $\v\al$ in the $\ka$ dimensional unit cube, where $\ka$ is the
dimension of $K$ and $\v\al$ satisfy the quantization condition
\begin{equation*}
 K\v\al=\text{integer vectors} .
\end{equation*}
$\v \al$'s that differ by an integer vector correspond to the same state.

Assume that the torus is a square of size 1: $ (X_1,X_2) \sim (X_1+1,X_2) \sim
(X_1,X_2+1)$ with $N_\phi$ magnetic flux quanta.  To simplify our result, we
will assume $N_\phi=$ even (see \eqn{NphiE}).  The Hamiltonian commutes with
$T^{(N)}_1$ and $T^{(N)}_2$, where $T^{(N)}_1$ generates translation
$(X_1,X_2)\to (X_1+\frac{1}{N_\phi},X_2)$ and $T^{(N)}_2$ generates translation
$(X_1,X_2)\to (X_1,X_2+\frac{1}{N_\phi})$.  $T^{(N)}_1$ and $T^{(N)}_2$ satisfy
the Heisenberg algebra
\begin{equation*}
 T^{(N)}_1T^{(N)}_2=\e^{\imth 2\pi \frac{N}{N_\phi}} T^{(N)}_2T^{(N)}_1
\end{equation*}
where $N$ is the total number of electrons.
The degenerate ground states $|\v\al\>$ form a representation of the
Heisenberg algebra:
for bosonic electrons,
\begin{align}
\label{T12alB1}
T^{(N)}_1 |\v\al\> &=\e^{\imth 2\pi \v q^T\v\al}|\v\al\>, 
\nonumber\\
T^{(N)}_2 |\v\al\> &=|\v\al+K^{-1}\v q\>, 
\end{align}
and for fermionic electrons,
\begin{align}
\label{T12alF1}
T^{(N)}_1 |\v\al\> &=(-)^{\v\ka^T\v q} \e^{\imth 2\pi \v q^T\v\al}|\v\al\>, 
\nonumber\\
T^{(N)}_2 |\v\al\> &=(-)^{\v\ka^T\v q}|\v\al+K^{-1}\v q\>.
\end{align}

For generic inverse-mass-matrix $g$ in \eqn{mtau}, the degenerate ground
states have a $\tau$ dependence $|\v\al;\tau\>$.
By deforming $\tau$ to $\tau'=\frac{b+a\tau}{d+c\tau}$, we can obtain
a non-Abelian geometric phase $U(\tau\to\tau')$.
$U(\tau\to\tau')$'s form a projective representation
of the modular group.

For two generators of the modular group
$S=U(\tau\to -1/\tau)=U(M_S)$ and
$T=U(\tau\to \tau+1)=U(M_T)$, we find that,
for bosonic electrons 
(see \eqn{bST}) 
\begin{align}
\label{STKB1}
 T_{\v\al\v\bt}&=
\e^{\frac{\imth\pi}{12}  (\v N^TK \v N-\v \ka^T\v N )}
\e^{\imth \pi (\v\al^TK\v\al-\frac{1}{12}\ka)}\del_{\v\al\v\bt}
\nonumber\\
 S_{\v\al\v\bt}&=
(-)^{(\v N^TK \v N-\v \ka^T\v N )/2}
\sum_{\v\bt\in UC}
\frac{\e^{ -\imth 2\pi \v\bt^T K\v\al }}
{ \sqrt{\det(K)} }
\end{align}
and for fermionic electrons
(see \eqn{fST}) 
\begin{align}
\label{STKF1}
 T_{\v\al\v\bt}&=
\e^{\frac{1}{12}\imth\pi  (\v N^TK \v N-\v \ka^T\v N -\ka)}\times
\nonumber\\
&\ \ \ \
\e^{\imth \pi (\v\al+\frac12 K^{-1}\v\ka)^TK(\v\al+\frac12 K^{-1}\v\ka)}\del_{\v\al\v\bt}
\nonumber\\
 S_{\v\al\v\bt}&=
(-)^{(\v N^TK \v N-\v \ka^T\v N )/2} \times
\nonumber\\
&\ \ \ \
\e^{ \frac12 \imth \pi \v\ka^T K^{-1} \v\ka }
\frac{ \e^{ -\imth 2\pi ( \v\al^T K\v\bt +\v\bt^T \v\ka ) } } { \sqrt{\det(K)} }
\end{align}
where $\v \ka^T=(K^{11},K^{22},\cdots,K^{\ka\ka})$ is two times the spin vector
of the Abelian FQH state.\cite{Wtoprev} We see that the $S$-matrix contains
information about the mutual statistics $2\pi \v \al^T K\v\bt$ between the
quasi-particles labeled by $\v\al$ and $\v\bt$, while the $T$-matrix contains
information about the statistics $\pi \v \al^T K\v\al$ of the quasi-particle
labeled by $\v\al$.\cite{Wrig,KW9327}

We have calculated the above unitary matrices $S,T$ carefully such that the
over all $U(1)$ phase is meanful. We have achieve this by choosing the
degenerate ground state wave functions on torus to be (1) holomorphic functions
of $\tau$ and (2) modular forms of level zero (the $\tau\to -1/\tau$ modular
transformation do not have the $\sqrt{-\imth \tau}$ factor). (1) is achieved by
choosing proper gauge for the background magnetic field, and (2) is achieved by
including proper powers of Dedekind $\eta(\tau)$ function in the ground state
wave functions (see \eqn{falcB} and \eqn{PsiKal}).  So the amibguity of the
ground state wave function is only a constant factor that is independent of
$\tau$.  Such a constant factor is a unitary matrix $W$ that is independent of
$\tau$.  So the ambiguity of $T,S$ is
\begin{align}
 T\to WTW^\dag,\ \ \ \
 S\to WSW^\dag.
\end{align} 

We note that $M_S$ generates a 90$^\circ$ rotation (see \eqn{XpXV}).  Also when
$\tau=\imth$, the torus has a 90$^\circ$ symmetry.  So $S=R_{90^\circ}$ is the
generator of the 90$^\circ$ rotation among the degenerate ground states when
$\tau=\imth$.  We also note that $M_SM_T= \bpm 0&-1\\1&1\epm$ and $(M_SM_T)^3=
\bpm -1&0\\0&-1\epm$ which is a 180$^\circ$ rotation (see \eqn{XpXV}).  So
$M_SM_T$ generates a 60$^\circ$ rotation.  When $\tau=\e^{\imth \pi/3}$, the
torus has a 60$^\circ$ symmetry.  So $U(M_SM_T)=U(M_S)U(M_T)=ST=R_{60^\circ}$
is the generator of the 60$^\circ$ rotation among the degenerate ground states
when $\tau=\e^{\imth\pi/3}$.
We find that, for bosonic electrons
\begin{align}
R_{60^\circ}&=
\e^{\frac{7}{12}\imth\pi  (\v N^TK \v N-\v \ka^T\v N)}
\e^{-\frac{1}{12}\imth\pi  \ka}
\frac{ 
\e^{\imth \pi (\v\bt-2\v\al)^TK\v\bt}
} { \sqrt{\det(K)} }
\end{align}
and for fermionic electrons
\begin{align}
 R_{60^\circ}&=
\e^{\frac{7}{12}\imth\pi  (\v N^TK \v N-\v \ka^T\v N)}
\e^{-\frac{1}{12}\imth\pi  \ka}
\e^{ \frac12 \imth \pi \v\ka^T K^{-1} \v\ka }
\times
\nonumber\\
&\ \ \ \ 
\frac{ \e^{ -\imth 2\pi ( \v\al^T K\v\bt +\v\bt^T \v\ka ) } } { \sqrt{\det(K)} }
\e^{\imth \pi (\v\bt+\frac12 K^{-1}\v\ka)^TK(\v\bt+\frac12 K^{-1}\v\ka)}
\nonumber\\
&=
\e^{\frac{7}{12}\imth\pi  (\v N^TK \v N-\v \ka^T\v N)}
\e^{-\frac{1}{12}\imth\pi  \ka}
\e^{ \frac34 \imth \pi \v\ka^T K^{-1} \v\ka }
\times
\nonumber\\
&\ \ \ \ 
\frac{ 
\e^{\imth \pi (\v\bt-2\v\al-K^{-1}\v\ka)^TK\v\bt}
} { \sqrt{\det(K)} }
\end{align}
We have $R_{60^\circ}^6=R_{90^\circ}^4=1$.

In $S,T$, we have a term that depend on $\v N^TK \v N-\v\ka^T \v N$.  Such a
term comes from the local relative motion (the $f_r$ part in \eqn{PsiKth}) and
is a pure $U(1)$ phase.  Other parts of $T,S$ come from the global
center-of-mass motion (the $f_c$ part in \eqn{PsiKth}) and are robust against
any local perturbations of the Hamiltonian.  Since
$R_{60^\circ}^6=R_{90^\circ}^4=1$ and the $\v N^TK \v N-\v\ka^T \v N$ term is
only a pure $U(1)$ phase, we see that such a $U(1)$ phase is quantized and is
also robust against any local perturbations of the Hamiltonian.  Therefore, $S$
and $T$ are robust against any local perturbations of the Hamiltonian.

We note that when $N_I=0$ mod 24, the $U(1)$ phase from $\v N^TK \v N-\v\ka^T
\v N$ is zero (mod $2\pi$).  In the following, we will assume $N_I=0$ mod 24
and consider only the bosonic case.  In this case, $S,T$ have simple forms
where $T$ is diagional and $S$ contains a row and column of the same index
which contain only positive elements and the element at the intersection of the
row and the column is 1.  (For more details see
\Ref{ZGT1251,CV1223,ZMP1233,LW}.) We can make such a row and column to be the
first row and the first column.  In this case,
$T_{11}=T_{\v\al=0,\v\al=0}=\e^{-\imth \pi \Del c/12}$ where $\Del c=\ka$ is
the chiral central charge.  (The chiral central charge is $c-\bar c$ where $c$
is the central charge for the right moving edge excitations and $\bar c$ is
the central charge for the left moving edge excitations.) We see that the
non-Abelian geometric phases contain information of the chiral central charge
mod 24.  

We like to remark that if we move $\tau$ around a samll cotractible loop, the
induced non-Abelian geometric phases are only a pure $U(1)$ phase.  Such a pure
$U(1)$ phase is related to the orbital spin\cite{WZ9253} or the Hall
viscosity.\cite{ASZ9597,TV0705,R0908} 

Although $(K,\v q,\v s)$ are not directly measurable, the ground state
degeneracy and $(T^{(N)}_1, T^{(N)}_2, S,T)$ 
are directly measurable (at least in numerical calculations).  Those quantities
give us a physical characterization of 2D topological orders, including the
chiral central charge mod 24.

This research is supported by NSF Grant No. DMR-1005541, NSFC 11074140, and
NSFC 11274192.  Research at Perimeter Institute is supported by the Government
of Canada through Industry Canada and by the Province of Ontario through the
Ministry of Research.


%

\appendix

\section{Theta functions}

The theta function
in \eqn{thNphi} is defined as
\begin{align*}
  \theta_{\al} (z \mid \tau) 
= \sum_{n\in \Z} \exp \{ \imth\pi \tau (n + \al)^2 + \imth 2 \pi (n + \al) z \}.
\end{align*}
$\theta_{\al} (z \mid \tau)$ has one zero on the torus
defined by $z\sim z+1$ and $z\sim z+\tau$.
$\th_\al(z|\tau)$ satisfies
\begin{align}
\label{qpcth}
&\ \ \ \ \th_\al(z+a+b\tau|\tau) =
\sum_{n} \e^{ \imth\pi \tau (n + \al)^2 + \imth 2 \pi (n + \al) (z+a+b\tau) }
\nonumber\\
&=
\e^{-\imth\pi\tau b^2 +\imth 2 \pi \al a}
\sum_{n} \e^{ \imth\pi \tau (n + \al+b)^2 + \imth 2 \pi (n + \al) z }
\nonumber\\
&= \e^{\imth 2 \pi \al a}\th_\al(z|\tau) \e^{-\imth\tau \pi b^2 - \imth 2\pi z b},
\ \ \ \ \
a,b=\text{integers} .
\end{align}
In particular, we find
\begin{align}
\label{thlmab}
\th_{l/m}(m(z+a+b\tau)|m\tau) = \th_{l/m}(mz|m\tau) 
\e^{-\imth m\tau \pi b^2 - \imth 2\pi z b m}  ,
\end{align}
and
\begin{align}
\label{thlmT1T2}
\th_{l/m}(m(z+\frac{1}{m})|m\tau) &= \th_{l/m}(mz|m\tau) \e^{\imth 2\pi l/m}  ,
\nonumber\\
\th_{l/m}(m(z+\frac{\tau}{m})|m\tau) &= 
\th_{(l+1)/m}(mz|m\tau)  \e^{-\imth \tau \pi/m  - \imth 2\pi z }.
\end{align}

Let us introduce four variants of the theta function:
\begin{align*}
 \th_{00}(z|\tau)& = \th_0(z|\tau), 
\nonumber\\
 \th_{10}(z|\tau)& = \e^{\frac14 \pi \imth\tau + \pi\imth z}
\th_0(z+ \frac12 \tau|\tau), 
\nonumber\\
 \th_{01}(z|\tau)& = \th_0(z+\frac12|\tau), 
\nonumber\\
 \th_{11}(z|\tau)& = \e^{\frac14 \pi \imth\tau + \pi\imth (z+\frac12)}
\th_0(z+\frac12+ \frac12 \tau|\tau). 
\end{align*}
$\th_{00}$,
$\th_{01}$, and
$\th_{10}$ are even and
$\th_{11}$ is odd in $z$:
\begin{align*}
 \th_{00}(z|\tau)& = \th_{00}(-z|\tau), 
&
 \th_{01}(z|\tau)& = \th_{01}(-z|\tau), 
\nonumber\\
 \th_{10}(z|\tau)& = \th_{10}(-z|\tau), 
&
 \th_{11}(z|\tau)& = -\th_{11}(-z|\tau). 
\end{align*}

From \eqn{qpcth}, we find that
\begin{align}
\label{qpcth11}
&\ \ \ \ 
\th_{11}(z+a+b\tau|\tau)
\\
&=
\th_{11}(z|\tau)
\e^{\imth\pi(a+b\tau)}
\e^{-\imth\tau \pi b^2 - \imth 2\pi (z+\frac12+\frac12 \tau) b},
\nonumber\\
&=(-)^{a+b}
\th_{11}(z|\tau) 
\e^{-\imth\tau \pi b^2 - \imth 2\pi z b},\ \ \ \ \
a,b=\text{integers} .
\nonumber 
\end{align}

We also have
\begin{align}
\label{taup1}
 \th_{11}(z|\tau+1)
& = \e^{\frac14 \imth \pi} \th_{11}(z|\tau)
\\
\label{invtau}
 \th_{11}(\frac{z}{\tau}|-\frac{1}{\tau})
& =-(-\imth \tau)^{1/2} \e^{\frac{\pi}{\tau}\imth z^2} \th_{11}(z|\tau)
\end{align}

\section{Multi-dimension theta functions}

The Riemann theta function is given by
\begin{equation}
\Th(\v z|\Om)
=\sum_{\v n\in Z^\ka} \e^{ \imth\pi \v n^T\Om \v n + \imth 2  \pi \v n^T \v z},
\end{equation}
for any symmetric complex matrix $\Om$ whose imaginary part is
positive definite.
It has the following properties:\cite{Mum83}
\begin{equation}
\Th(\v z+\v a +\Om \v b|\Om)
=\e^{ -\imth \pi \v b^T\Om \v b - \imth 2\pi \v b^T \v z}
\Th(\v z|\Om) ,
\end{equation}
for any integer vectors $\v a$ and $\v b$.

We also have
\begin{equation}
\Th(\v z|\Om+J) =
\Th(\v z+\frac{\text{diag} (J)}{2}|\Om) ,
\end{equation}
for any symmetric integer matrix $J$, and
\begin{equation}
\Th(\Om^{-1}\v z|-\Om^{-1}) =
\sqrt{\det(-\imth \Om)} \e^{\imth\pi \v z^T\Om^{-1}\v z}\Th(\v z|\Om) .
\end{equation}
These two relation lead to\cite{Mum83}
\begin{align}
&\ \ \ \Th([(C\Om+D)^T]^{-1} \v z| (A\Om+B)(C\Om+D)^{-1})
\nonumber\\
&=
\zeta_\ga \sqrt{ \text{det}(C\Om+D) } \e^{\imth\pi \v z^T(C\Om+D)^{-1} C\v z}
\Th(\v z|\Om),
\end{align}
where 
\begin{align}
& \ga=\bpm A & B\\
          C & D\epm \in Sp(2\ka,\Z),
\nonumber\\
&
\text{diag}(A^T C), \text{diag}(B^T D) = \text{even vectors} .
\end{align}
$\zeta_\ga$ is an $\ga$ dependent phase satisfying $\zeta_\ga^8=1$.

Let
\begin{equation*}
F_K(\v Z|\tau)= \Th(K\v Z|K\tau)  .
\end{equation*}
We find that
\begin{align*}
 F_K(\v Z+\v \al+\v b\tau|\tau) &= \Th(K(\v Z+\v \al+\v b\tau)|K\tau)
\nonumber\\
&=
\e^{-\imth \pi \tau \v b^T K\v b - \imth 2\pi \v b^TK\v Z}
F_K(\v Z|\tau),
\nonumber\\
\text{for } & K\v\al, \v b = \text{integer vectors},
\end{align*}
\begin{equation*}
F_K(\v Z|\tau+1) =
F_K(\v Z+\frac{K^{-1}\v\ka}{2}|\tau) ,
\end{equation*}
and
\begin{align*}
&\ \ \ \  F_K(\v Z|-1/\tau)
= \Th(K\v Z|-K/\tau) 
\nonumber\\
&= 
\sqrt{\det(-\imth K^{-1}\tau)} \e^{\imth \pi \tau\v Z^T K \v Z} 
\Th(\tau \v Z|K^{-1}\tau)
\nonumber\\
&= \sqrt{\det(-\imth K^{-1}\tau)} \e^{\imth \pi \tau\v Z^T K \v Z}
\sum_{\v n\in Z^\ka} \e^{\tau\imth\pi \v n^TK^{-1} \v n+\imth 2\pi \tau \v n^T\v Z}
\nonumber\\
&= 
\sqrt{\det(-\imth K^{-1}\tau)} \e^{\imth \pi \tau\v Z^T K \v Z}
\times
\nonumber\\
&\ \ \ \ \ \
\sum_{\v\al\in UC} \sum_{\v n\in Z^\ka} 
\e^{\tau\imth\pi (\v n+\v\al)^TK (\v n+\v\al)+\imth 2\pi(\v n+\v\al)^TK\v Z\tau}
\nonumber\\
&= 
\sqrt{\det(-\imth K^{-1}\tau)} \e^{\imth \pi \tau\v Z^T K \v Z}
\times
\nonumber\\
&\ \ \ \ \ \
\sum_{\v\al\in UC} 
\e^{\tau\imth\pi \v\al^TK \v\al+\imth 2\pi\v\al^TK\v Z\tau}
F_K(\tau(\v Z+\v\al)|\tau)
\nonumber\\
&= 
\sum_{\v\al\in UC} 
\sqrt{\det(-\imth K^{-1}\tau)} 
\e^{\tau\imth\pi (\v Z+\v\al)^TK (\v Z+\v\al)}
F_K(\tau(\v Z+\v\al)|\tau)
\end{align*}

Let us define $f^{(\v\al,\v \eta)}(\v Z|\tau)$ as
\begin{align*}
&\ \ \ \ f^{(\v\al,\v \eta)}(\v Z|\tau)
\nonumber\\
& =\sum_{\v n\in Z^\ka} \e^{ 
\imth\pi (\v n + \v\al+\v\eta)^TK\tau (\v n + \v\al+\v\eta)
+ \imth 2  \pi (\v n +\v\al+\v\eta)^T K (\v Z -\v  \eta)},
\nonumber\\
& =
\e^{\imth \pi(\v\al+\v\eta)^TK\tau(\v\al+\v\eta)
+\imth 2\pi(\v\al+\v\eta)^TK(\v Z- \v\eta)}\times
\nonumber\\
&\ \ \ \ \ \ \ \ \
\sum_{\v n\in Z^\ka} \e^{ 
\imth\pi \v n^TK\tau \v n 
+ \imth 2  \pi \v n^T K [\v Z +(\v\al+\v\eta)\tau-\v  \eta]}
\nonumber\\
&= \e^{\imth \pi(\v\al+\v\eta)^TK\tau(\v\al+\v\eta)
+\imth 2\pi(\v\al+\v\eta)^TK(\v Z- \eta)}
\times
\nonumber\\
&\ \ \ \ \ \ \ \ \
\Th(K[\v Z+(\v\al+\v\eta)\tau-\v \eta]|K\tau)  .
\nonumber\\
&= \e^{\imth \pi(\v\al+\v\eta)^TK\tau(\v\al+\v\eta)
+\imth 2\pi(\v\al+\v\eta)^TK(\v Z- \eta)} \times
\nonumber\\
&\ \ \ \ \ \ \ \ \
F_K(\v Z+(\v\al+\v\eta)\tau-\v \eta|\tau)  .
\end{align*}
It has the following properties:
For any integer vectors $\v a$ and $\v b$, we have
\begin{align}
\label{faletaab}
&\ \ \ \
f^{(\v\al,\v \eta)}(\v Z+\v a +\v b \tau|\tau),
\nonumber\\
&= \e^{\imth 2\pi\v\eta^TK (\v a+\v b)}
\e^{-\imth \pi \tau \v b^T K\v b - \imth 2\pi \v b^TK \v Z}
f^{(\v\al,\v \eta)}(\v Z|\tau) ,
\end{align}
\begin{align}
\label{faletatrans}
&\ \ \
f^{(\v\al,\v \eta)}(\v Z+K^{-1}\v a |\tau)
\nonumber\\
&= \e^{\imth 2\pi(\v\al+\v\eta)^T\v a} 
f^{(\v\al,\v \eta)}(\v Z|\tau) ,
\\
&\ \ \
f^{(\v\al,\v \eta)}(\v Z+K^{-1}\v b \tau |\tau)
\nonumber\\
&= \e^{\imth 2\pi\v\eta^T\v b}
\e^{-\imth \pi \tau \v b^T K^{-1}\v b - \imth 2\pi \v b^T\v Z}
f^{(\v\al+K^{-1}\v b,\v \eta)}(\v Z|\tau) ,
\nonumber 
\end{align}
and
\begin{align}
\label{faletatau1}
&\ \ \ f^{(\v\al,\v\eta)}(\v Z|\tau+1)
\nonumber\\
&= \e^{\imth \pi(\v\al+\v\eta)^TK(\tau+1)(\v\al+\v\eta)
+\imth 2\pi(\v\al+\v\eta)^TK(\v Z-\eta)}\times
\nonumber\\
&\ \ \ \ \ \ \ \ \ \ \ \
F_K(\v Z+(\v\al+\v\eta)(\tau+1)-\v\eta|\tau+1)  
\nonumber\\
&= \e^{\imth \pi(\v\al+\v\eta)^TK(\tau+1)(\v\al+\v\eta)
+\imth 2\pi(\v\al+\v\eta)^TK(\v Z-\eta)}\times
\nonumber\\
&\ \ \ \ \ \ \ \ \ \ \ \
F_K(\v Z+(\v\al+\v\eta)\tau+\v\al+\frac{K^{-1}\v\ka }{2}|\tau)  
\nonumber\\
&= \e^{\imth \pi(\v\al+\v\eta)^TK(\tau+1)(\v\al+\v\eta)
+\imth 2\pi(\v\al+\v\eta)^TK(\v Z-\eta)}\times
\nonumber\\
&\ \ \ \ \ \ \ \ \ \ \ \
F_K(\v Z+(\v\al+\v\eta)\tau+\frac{K^{-1}\v\ka }{2}|\tau)  
.
\end{align}
where we have used $F_K(\v Z|\tau)=F_K(\v Z+\v\al|\tau)$.
Here $\v \ka=(K_{11},K_{22}, ...)$.

If $\v\eta=\v 0$ and $\v \ka/2$ is an integer vector, we have
\begin{align}
\label{falBtau1}
f^{(\v\al,\v 0)}(\v Z|\tau+1)
&= \e^{\imth \pi\v\al^TK(\tau+1)\v\al
+\imth 2\pi\v\al^TK\v Z}
F_K(\v Z+\v\al\tau |\tau)
\nonumber\\
&= \e^{\imth \pi\v\al^TK\v\al}
f^{(\v\al,\v 0)}(\v Z|\tau)
.
\end{align}
If $\v\eta=K^{-1}\v\ka/2$, we have
\begin{align*}
&\ \ \ f^{(\v\al,\v\eta)}(\v Z|\tau+1)
\nonumber\\
&= \e^{\imth \pi(\v\al+\v\eta)^TK(\tau+1)(\v\al+\v\eta)
+\imth 2\pi(\v\al+\v\eta)^TK(\v Z-\eta)}\times
\nonumber\\
&\ \ \ \ \ \ \ \
F_K(\v Z+(\v\al+\v\eta)\tau -\v\eta|\tau)
,
\end{align*}
or
\begin{align}
\label{falFtau1}
&\ \ \ 
f^{(\v\al,K^{-1}\v\ka/2)}(\v Z|\tau+1)
\nonumber\\
&= \e^{\imth \pi(K\v\al+\frac12\v\ka)^TK^{-1}(K\v\al+\frac12\v\ka) }
f^{(\v\al,K^{-1}\v\ka/2)}(\v Z|\tau)
.
\end{align}

We also have
\begin{align}
\label{faletatauinv}
&\ \ \ \ 
f^{(\v\al,\v\eta)}(\v Z|-1/\tau)
\nonumber\\
&= \e^{-\tau^{-1}\imth \pi(\v\al+\v\eta)^TK(\v\al+\v\eta)
+\imth 2\pi(\v\al+\v\eta)^TK(\v Z-\eta)}\times
\nonumber\\
&\ \ \ \ \ \ \ \
F_K(\v Z-(\v\al+\v\eta)\tau^{-1}-\v\eta|-1/\tau)  
\nonumber\\
&= \e^{-\tau^{-1}\imth \pi(\v\al+\v\eta)^TK(\v\al+\v\eta)
+\imth 2\pi(\v\al+\v\eta)^TK(\v Z-\eta)}\times
\nonumber\\
& \ \ \ \ \ \
\sum_{\v\bt\in UC} \sqrt{\det(-\imth K^{-1}\tau)} 
\times
\nonumber\\
& \ \ \ \ \ \
\e^{\tau\imth\pi (\v Z-(\v\al+\v\eta)\tau^{-1}-\v\eta+\v\bt)^TK 
(\v Z-(\v\al+\v\eta)\tau^{-1}-\v\eta+\v\bt)}\times
\nonumber\\
&\ \ \ \ \ \ \ \ \ \ \ \ \ \ \ \
F_K(\tau(\v Z-(\v\al+\v\eta)\tau^{-1}-\v\eta+\v\bt)|\tau) \times
\nonumber\\
&=\sum_{\v\bt\in UC} \sqrt{\det(-\imth K^{-1}\tau)} 
\times
\nonumber\\
& \ \ \ \ \ \
\e^{\tau\imth\pi (\v Z+\v\bt-\v\eta)^TK (\v Z+\v\bt-\v\eta)
-\imth 2\pi \v\bt^T K(\v\al+\v\eta) } \times
\nonumber\\
&\ \ \ \ \ \ \ \ \ \ \ \ \ \ \ \
F_K(\tau\v Z-(\v\al+\v\eta)-\tau\v\eta+\tau\v\bt|\tau)
\nonumber\\
&=\sum_{\v\bt\in UC} \sqrt{\det(-\imth K^{-1}\tau)} 
\times
\nonumber\\
& \ \ \ \ \ \
\e^{\tau\imth\pi (\v Z+\v\bt-\v\eta)^TK (\v Z+\v\bt-\v\eta)
-\imth 2\pi \v\bt^T K(\v\al+\v\eta) }\times
\nonumber\\
&\ \ \ \ \ \ \ \ \ \ \ \ \ \ \ \
F_K(\tau\v Z+\tau(\v\bt-\v\eta)-\v\eta|\tau)
.
\end{align}
where we have used $F_K(\v Z|\tau)=F_K(\v Z+\v\al|\tau)$.

When $\v \eta=\v 0$, we find
\begin{align*}
&\ \ \ \ 
f^{(\v\al,\v 0)}(\v Z|-1/\tau)
\nonumber\\
&=\sum_{\v\bt\in UC} \sqrt{\det(-\imth K^{-1}\tau)} 
\e^{\tau\imth\pi (\v Z+\v\bt)^TK (\v Z+\v\bt)
-\imth 2\pi \v\bt^T K\v\al }
\times
\nonumber\\
&\ \ \ \ \ \ 
F_K(\tau\v Z+\tau\v\bt|\tau)
\nonumber\\
&= \sqrt{\det(-\imth K^{-1}\tau)} \e^{\tau\imth\pi \v Z^TK \v Z }
\sum_{\v\bt\in UC} \e^{ -\imth 2\pi \v\bt^T K\v\al }
f^{(\v\bt,\v 0)}(\tau\v Z|\tau)
.
\end{align*}
or
\begin{align}
\label{falBtauinv}
&\ \ \ \ 
f^{(\v\al,\v 0)}(\v Z/\tau|-1/\tau)
\\
&=
(-\imth\tau)^{\ka/2}
\e^{\imth\pi \v Z^TK \v Z/\tau }
\sum_{\v\bt\in UC}
\frac{\e^{ -\imth 2\pi \v\bt^T K\v\al }}
{ \sqrt{\det(K)} }
f^{(\v\bt,\v 0)}(\v Z|\tau)
.
\nonumber 
\end{align}
Note that
\begin{equation*}
 \sum_{\v\ga\in UC}
\frac{\e^{ -\imth 2\pi \v\al^T K\v\ga }} { \sqrt{\det(K)} }
\frac{\e^{ +\imth 2\pi \v\bt^T K\v\ga }} { \sqrt{\det(K)} }=
\del_{\v\al,\v\bt} .
\end{equation*}
\Eqn{falBtauinv} can also be rewritten as
\begin{align}
\label{falBtauinv1}
f^{(\v\al,\v 0)}(\v Z|\tau)
&=
(-\imth\tau)^{-\ka/2}
\e^{-\imth\pi \v Z^TK \v Z/\tau } \times
\\
&\ \ \ \ \ \ \
\sum_{\v\bt\in UC}
\frac{\e^{ \imth 2\pi \v\bt^T K\v\al }}
{ \sqrt{\det(K)} }
f^{(\v\bt,\v 0)}(\v Z/\tau|-1/\tau)
.
\nonumber 
\end{align}

When $\v \eta=K^{-1}\v \ka/2$, we note that
$F_K(\tau\v Z+\tau(\v\bt-\v\eta)-\v\eta|\tau)=
F_K(\tau\v Z+\tau(\v\bt-\v\eta)+\v\eta|\tau)$.
Thus
\begin{align*}
&\ \ \ \ 
f^{(\v\al,\v\eta)}(\v Z|-1/\tau)
\nonumber\\
&=\sum_{\v\bt\in UC} \sqrt{\det(-\imth K^{-1}\tau)} 
\times
\nonumber\\
&\ \ \ \ \ \ \
\e^{\tau\imth\pi (\v Z+\v\bt-\v\eta)^TK (\v Z+\v\bt-\v\eta)
-\imth 2\pi \v\bt^T K(\v\al+\v\eta) }\times
\nonumber\\
&\ \ \ \ \ \ \ \ \ \ \ \ \ \ \ \
F_K(\tau\v Z+\tau(\v\bt-\v\eta)+\v\eta|\tau)
\nonumber\\
&= \sqrt{\det(-\imth K^{-1}\tau)} \e^{\tau\imth\pi \v Z^TK \v Z }
\times
\nonumber\\
&\ \ \ \ \ \ \
\sum_{\v\bt\in UC} 
\e^{ -\imth 2\pi ( \v\bt^T K\v\al +2\v\bt^T K\v\eta -\v\eta^T K\v\eta ) }
f^{(\v\bt,\v \eta)}(\tau\v Z|\tau)
,
\end{align*}
or
\begin{align}
\label{falFtauinv}
&\ \ \ 
f^{(\v\al,K^{-1}\v\ka/2)}  (\v Z/\tau|-1/\tau)
\nonumber\\
&= (-\imth \tau)^{\ka/2} \e^{\imth\pi \v Z^TK \v Z/\tau }
\e^{ \imth \pi \v\ka^T K^{-1} \v\ka/2 }\times
\nonumber\\
&
\sum_{\v\bt\in UC} 
\frac{ \e^{ -\imth 2\pi ( \v\bt^T K\v\al +\v\bt^T \v\ka ) } } { \sqrt{\det(K)} }
f^{(\v\bt,K^{-1}\v \ka/2)}(\v Z|\tau)
,
\end{align}
which can also be rewritten as
\begin{align}
\label{falFtauinv1}
&\ \ \ 
f^{(\v\al,K^{-1}\v\ka/2)}  (\v Z|\tau)
\\
&= (-\imth \tau)^{-\ka/2} \e^{-\imth\pi \v Z^TK \v Z/\tau }
\e^{-\imth \pi \v\ka^T K^{-1} \v\ka/2 }\times
\nonumber\\
&
\sum_{\v\bt\in UC} 
\frac{ \e^{ \imth 2\pi ( \v\al^T K\v\bt +\v\al^T \v\ka ) } } { \sqrt{\det(K)} }
f^{(\v\bt,K^{-1}\v \ka/2)}(\v Z/\tau|-1/\tau)
.
\nonumber 
\end{align}

From \eqn{falBtauinv1}, we find that
\begin{align}
\label{falB60}
&\ \ \ \ 
f^{(\v\al,\v 0)}(\tau\v Z|\tau)
\\
&=
(-\imth\tau)^{-\ka/2} \e^{-\imth\pi \tau \v Z^TK \v Z }
\sum_{\v\bt\in UC} \frac{\e^{ \imth 2\pi \v\bt^T K\v\al }} { \sqrt{\det(K)} }
f^{(\v\bt,\v 0)}(\v Z|-1/\tau)
\nonumber\\
&=
(-\imth\tau)^{-\ka/2} \e^{-\imth\pi \tau \v Z^TK \v Z }
\times
\nonumber\\
&\ \ \ \ \ \ \
\sum_{\v\bt\in UC} \frac{\e^{ \imth 2\pi \v\bt^T K\v\al }} { \sqrt{\det(K)} }
\e^{-\imth \pi \v\bt^TK\v\bt}
f^{(\v\bt,\v 0)}(\v Z|-\frac{1}{\tau}+1)
.
\nonumber 
\end{align}

From \eqn{falFtauinv1}, we find that
\begin{align}
\label{falF60}
&\ \ \ \ 
f^{(\v\al,K^{-1}\v\ka/2)}  (\tau\v Z|\tau)
\\
&= (-\imth \tau)^{-\ka/2} \e^{-\imth\pi \tau \v Z^TK \v Z }
\e^{-\imth \pi \v\ka^T K^{-1} \v\ka/2 }\times
\nonumber\\
&\ \ \ \ \ \ \ \ \ \ \
\sum_{\v\bt\in UC} 
\frac{ \e^{ \imth 2\pi ( \v\al^T K\v\bt +\v\al^T \v\ka ) } } { \sqrt{\det(K)} }
f^{(\v\bt,K^{-1}\v \ka/2)}(\v Z|-1/\tau)
\nonumber\\
&= (-\imth \tau)^{-\ka/2} \e^{-\imth\pi \tau \v Z^TK \v Z }
\e^{-\imth \pi \v\ka^T K^{-1} \v\ka/2 }
\times
\nonumber\\
&\ \ \ \ \ \ \
\sum_{\v\bt\in UC} 
\frac{ \e^{ \imth 2\pi ( \v\al^T K\v\bt +\v\al^T \v\ka ) } } { \sqrt{\det(K)} }
\times
\nonumber\\
&\ \ \
\e^{-\imth \pi(K\v\bt+\frac12\v\ka)^TK^{-1}(K\v\bt+\frac12\v\ka) }
f^{(\v\bt,K^{-1}\v \ka/2)}(\v Z|-1/\tau+1)
.
\nonumber 
\end{align}

\section{Algebra of modular transformations and translations}

The translation $T_{\v d}$ and modular transformation $U(M)$ all act within
the space of degenerate ground states. There is an algebraic relation between
those operators. From \eqn{UPsi}, we see that
\begin{align*}
&\ \ \ \
U(M)T_{\v d} \Phi_\ga(\{\v X_i\}|\tau) = U(M) \Phi_\ga(\{\v X_i+\v d\}|\tau) 
\nonumber \\
&=U_G \Phi_\ga(\{M^{-1}\v X_i+\v d\}|\tau') 
\nonumber\\
&=U_G \Phi_\ga(\{M^{-1}(\v X_i+M\v d)\}|\tau') .
\end{align*}
Therefore
\begin{equation}
\label{UTTU}
 \e^{\imth \th} U(M)T_{\v d}=T_{M\v d}U(M)  .
\end{equation}

Let us determine the possible phase factor $\e^{\imth \th}$ for some special
cases.
Consider the modular transformation $\tau\to \tau'=\tau+1$ generated by
\begin{equation*}
 M_T=\bpm
1 & 1\\
0 & 1
\epm,\ \ \ \ \ 
 M_T^{-1}=\bpm
1 & -1\\
0 & 1
\epm .
\end{equation*}
We first calculate
\begin{align*}
&\ \ \ \
U(M_T)T_1 \Phi_\ga(\{\v X_i\}|\tau) = U(M_T) \Phi_\ga(\{\v X_i+\v d_1\}|\tau) 
\nonumber \\
&= 
\e^{-\imth  \pi N_\phi \sum_i (X_i^2)^2 }  
\Phi_\ga(\{M_T^{-1}\v X_i+\v d_1\}|\tau') 
.
\end{align*}
where $\v d_1=(\frac{1}{N_\phi},0)$.  We note that
\begin{equation*}
 M_T\v d_1=
\bpm
1 & 1\\
0 & 1
\epm
\bpm
\frac{1}{N_\phi} \\
0 
\epm
=\v d_1 .
\end{equation*}
Thus we next calculate
\begin{align*}
&\ \ \ \
T_1 U(M_T) \Phi_\ga(\{\v X_i\}|\tau) 
\nonumber\\
&= 
T_1\e^{-\imth  \pi N_\phi \sum_i (X_i^2)^2 } \Phi_\ga(\{M_T^{-1}\v X_i\}|\tau') 
\nonumber \\
&= 
\e^{-\imth  \pi N_\phi \sum_i (X_i^2)^2 }  
\Phi_\ga(\{M_T^{-1}(\v X_i+\v d_1)\}|\tau') 
\nonumber\\
&= 
\e^{-\imth  \pi N_\phi \sum_i (X_i^2)^2 }  
\Phi_\ga(\{M_T^{-1}\v X_i+\v d_1\}|\tau') 
.
\end{align*}
Therefore
\begin{equation*}
 U(M_T)T_1=T_1U(M_T).
\end{equation*}

To obtain the algebra between $U(M_T)$ and $T_2$, we first
calculate
\begin{align*}
&\ \ \ \
U(M_T)T_2 \Phi_\ga(\{\v X_i\}|\tau) 
\nonumber\\
&= U(M_T) 
\e^{\imth 2\pi \sum_i X^1_i}
\Phi_\ga(\{\v X_i+\v d_2\}|\tau) 
\nonumber \\
&= 
\e^{-\imth  \pi N_\phi \sum_i (X_i^2)^2 }  
\e^{\imth 2\pi \sum_i (X^1_i-X^2_i)}
\Phi_\ga(\{M_T^{-1}\v X_i+\v d_2\}|\tau') 
,
\end{align*}
where $\v d_2=(0,\frac{1}{N_\phi})$.
We note that
\begin{equation*}
 M_T\v d_2=
\bpm
1 & 1\\
0 & 1
\epm
\bpm
0\\
\frac{1}{N_\phi} \\
\epm
=\v d_1 +\v d_2.
\end{equation*}
Thus we next calculate
\begin{align*}
&\ \ \ \
T_2T_1U(M_T) \Phi_\ga(\{\v X_i\}|\tau) 
\nonumber \\
&= 
T_2T_1\e^{-\imth  \pi N_\phi \sum_i (X_i^2)^2 }  
\Phi_\ga(\{M_T^{-1}\v X_i\}|\tau') 
\nonumber\\
&= 
T_2\e^{-\imth  \pi N_\phi \sum_i (X_i^2)^2 }  
\Phi_\ga(\{M_T^{-1}(\v X_i+\v d_1)\}|\tau') 
\nonumber\\
&= 
\e^{\imth 2\pi \sum_i X^1_i}
\e^{-\imth  \pi N_\phi \sum_i (X_i^2+\frac{1}{N_\phi})^2 }  \times
\nonumber\\
&\ \ \ \ \ \ \ \ \ \ \ 
\Phi_\ga(\{M_T^{-1}(\v X_i+\v d_1+\v d_2)\}|\tau') 
\nonumber\\
&= 
\e^{-\imth \pi N/N_\phi}
\e^{\imth 2\pi \sum_i (X^1-X^2_i)_i}
\e^{-\imth  \pi N_\phi \sum_i (X_i^2)^2 }  \times
\nonumber\\
&\ \ \ \ \ \ \ \ \ \ \ 
\Phi_\ga(\{M_T^{-1}\v X_i+\v d_2)\}|\tau') 
,
\end{align*}
We see that
\begin{equation*}
 U(M_T)T_2= 
\e^{\imth \pi \frac{n}{m}}
T_2T_1U(M_T).
\end{equation*}

Next we consider the modular transformation $\tau\to \tau'=-1/\tau$ 
generated by
\begin{equation*}
 M_S=\bpm
0 & -1\\
1 & 0
\epm,\ \ \ \ \ 
 M_S^{-1}=\bpm
0 & 1\\
-1 & 0
\epm .
\end{equation*}
From 
\begin{align*}
U(M_S) \Phi_\ga(\{\v X_i\}|\tau) 
&= 
\e^{-\imth  2 \pi N_\phi \sum_i X_i^1 X_i^2 }  
\Phi_\ga(\{M_S^{-1}\v X_i\}|\tau') 
,
\end{align*}
we find
\begin{align*}
&\ \ \ \ 
U(M_S) U(M_S) \Phi_\ga(\{\v X_i\}|\tau) 
\nonumber\\
&= U(M_S)
\e^{-\imth  2 \pi N_\phi \sum_i X_i^1 X_i^2 }  
\Phi_\ga(\{M_S^{-1}\v X_i\}|\tau') 
\nonumber\\
&=
\e^{-\imth  2 \pi N_\phi \sum_i X_i^1 X_i^2 }  
\e^{-\imth  2 \pi N_\phi \sum_i X_i^2 (-X_i^1) }  
\Phi_\ga(\{-\v X_i\}|\tau) 
\nonumber\\
&=
\Phi_\ga(\{-\v X_i\}|\tau) 
.
\end{align*}
We see that $U(M_S) U(M_S)=U(-1)$ generates to transformation $\v X\to -\v X$.
We can show that
\begin{equation}
\label{Um1T12}
U(-1) T_1 U^\dag(-1)=T^\dag_1,\ \ \ \ \ \ 
U(-1) T_2 U^\dag(-1)=T^\dag_2 .
\end{equation}
%
Clearly $C^2=1$.

Let us first calculate
\begin{align*}
&\ \ \ \ 
U(M_S) T_1 \Phi_\ga(\{\v X_i\}|\tau) 
\nonumber\\
&= U(M_S)
\Phi_\ga(\{M_S^{-1}\v X_i+\v d_1\}|\tau) 
\nonumber\\
&=
\e^{-\imth  2 \pi N_\phi \sum_i X_i^1 X_i^2 }  
\Phi_\ga(\{M_S^{-1}\v X_i+\v d_1\}|\tau') 
.
\end{align*}
Since $M_S\v d_1=\v d_2$, we next consider
\begin{align*}
&\ \ \ \ 
T_2 U(M_S)  \Phi_\ga(\{\v X_i\}|\tau) 
\nonumber\\
&=
T_2 \e^{-\imth  2 \pi N_\phi \sum_i X_i^1 X_i^2 }  
\Phi_\ga(\{M_S^{-1}\v X_i\}|\tau') 
\nonumber\\
&=
\e^{\imth 2\pi \sum_i X^1_i}
\e^{-\imth  2 \pi N_\phi \sum_i X_i^1 (X_i^2+\frac{1}{N_\phi}) } \times
\nonumber\\
&\ \ \ \ \ \ \ \ \ \ \ \ 
\Phi_\ga(\{M_S^{-1}(\v X_i+\v d_2)\}|\tau') 
\nonumber\\
&=
\e^{-\imth  2 \pi N_\phi \sum_i X_i^1 X_i^2 } 
\Phi_\ga(\{M_S^{-1}(\v X_i+\v d_2)\}|\tau') 
.
\end{align*}
We see that
\begin{align*}
 U(M_S)T_1&=T_2 U(M_S) ,
\nonumber\\
 U(M_S)T_2&=T^\dag_1 U(M_S) ,
\end{align*}
where we have used \eqn{Um1T12}.
Let us introduce
$ T=\e^{\imth \th} U(M_T)$ and
$ S= U(M_S)$,
where the value of $\th$ is chosen to simplify $T$.
We find that
\begin{align}
\label{TST1T2a}
 TT_1&=T_1T, &  TT_2&=\e^{\imth \pi\frac{n}{m}} T_2T_1 T,
\nonumber\\
 ST_1&=T_2S, &  ST_2&= T_1^\dag S.
\end{align}

%
%
%
%
%
%
%
%

\bibliography{../../bib/wencross,../../bib/all,../../bib/publst,./bib}

\end{document}